\documentclass[journal]{IEEEtran}
\usepackage{amsthm}
\theoremstyle{definition}

\usepackage{booktabs}
\usepackage{tabularx}
\usepackage{array}
\usepackage{subcaption}
\usepackage{svg}

\usepackage{tikz}
\usetikzlibrary{positioning,arrows.meta,decorations.pathreplacing,shadows.blur}
\usepackage{xurl} % better line breaks for monospace-like strings
\usepackage[ruled,vlined,linesnumbered]{algorithm2e}
\SetKwInput{KwIn}{Input}
\SetKwInput{KwOut}{Output}

\newcolumntype{L}[1]{>{\raggedright\arraybackslash}p{#1}}
\newcolumntype{Y}{>{\raggedright\arraybackslash}X}

\usepackage{amsmath}   % advanced math environments (equation, align, cases, etc.)
\usepackage{amssymb}   % additional math symbols
\usepackage{bm}        % bold math symbols (\bm{} for vectors/matrices)
\usepackage{microtype} % improves spacing and justification (recommended)
\usepackage{graphicx}  % for including figures
\usepackage{tikz}      % for diagrams (if you want to draw schema figures)
\usepackage{booktabs}  % professional tables
\usepackage{multirow}  % merge table rows
\usepackage{array}     % custom column formatting
\usepackage{hyperref}  % clickable references and links
\usepackage{cite}      % compressed, sorted citations [1]–[3]
\usepackage{float}
\usepackage{enumitem}
\usepackage[table]{xcolor}
\definecolor{cyanblue}{RGB}{224,238,255}
\usepackage{graphicx}
\usepackage{subcaption}
\graphicspath{{figures/}} % path to your images
\usepackage{tcolorbox}
\usepackage[ruled,vlined]{algorithm2e}
\usepackage{booktabs}
\usepackage{siunitx}
\usepackage{longtable}
\usetikzlibrary{positioning,calc,arrows.meta,fit,shapes,backgrounds}

\usepackage{tikz}
\usepackage[edges]{forest}
\usepackage{xcolor}
\usetikzlibrary{calc}

\usepackage{adjustbox}
\usepackage{caption}
\usepackage{amssymb}    % for \checkmark
\usepackage{pifont}     % for ding symbols
\usepackage{xcolor}     % for coloring if you like
 
\newcolumntype{Y}{S[table-format=1.3]}

\usepackage{bm}   % bold math symbols
\usepackage{listings}
\usepackage{xcolor} % for colors
\usepackage{listings}
\usepackage{xcolor}
\usepackage{xcolor}
\usepackage[T1]{fontenc}
\usepackage[utf8]{inputenc} % if using pdfLaTeX; XeLaTeX/LuaLaTeX don't need this
\usepackage{listings}
\usepackage{xcolor}
\usepackage[T1]{fontenc}
\usepackage[utf8]{inputenc} % if using pdfLaTeX

\lstdefinelanguage{json}{
  morestring=[b]",
  moredelim=[s][\color{black}]{\{}{\}},
  moredelim=[s][\color{black}]{[}{]},
  stringstyle=\color{brown},
  showstringspaces=false,
}

\lstdefinestyle{jsonschema}{
  language=json,
  basicstyle=\ttfamily\footnotesize,
  frame=single,
  breaklines=true,
  columns=fullflexible,
  keepspaces=true,
  showstringspaces=false,
}
\lstdefinestyle{plain}{
  basicstyle=\ttfamily\footnotesize,
  frame=single,
  breaklines=true,
  columns=fullflexible,
  keepspaces=true,
  showstringspaces=false,
}

\lstdefinestyle{python}{
  language=Python,
  basicstyle=\ttfamily\footnotesize,
  frame=single,
  breaklines=true,
  columns=fullflexible,
  keepspaces=true,
  showstringspaces=false,
  keywordstyle=\color{blue},
  commentstyle=\color{gray},
  stringstyle=\color{brown},
}
\ifCLASSINFOpdf
\else
\fi
\begin{document}
%\title{$\alpha^3$-Bench: A Dual-Domain 6G Benchmark for LLM-based Conversational Reasoning Agents in Autonomous UAV Systems}

\title{6G-Bench: An Open Benchmark for Semantic Communication and Network-Level Reasoning with Foundation Models in AI-Native 6G Networks}

\author{
\IEEEauthorblockN{
Mohamed~Amine~Ferrag\IEEEauthorrefmark{1}\IEEEauthorrefmark{3},
Abderrahmane~Lakas\IEEEauthorrefmark{1},
Merouane~Debbah\IEEEauthorrefmark{2}
}
\\
\IEEEauthorblockA{\IEEEauthorrefmark{1}
Department of Computer and Network Engineering,  
United Arab Emirates University, UAE
} \\
\IEEEauthorblockA{\IEEEauthorrefmark{2}
6G Research Center (6GRC), Khalifa University, UAE
}\\
\IEEEauthorblockA{\IEEEauthorrefmark{3}
Corresponding author: \texttt{mohamed.ferrag@uaeu.ac.ae}
}
}


%                     ^^^-Do not want these spaces!

\markboth{ }%
{Shell \MakeLowercase{\textit{et al.}}: Bare Demo of IEEEtran.cls for IEEE Journals}



\maketitle

%\begin{abstract}
%\end{abstract}

\begin{abstract}
Emerging sixth-generation (6G) networks are increasingly envisioned as AI-native, intent-driven systems in which foundation models act as high-level reasoning and coordination layers above standardized network functions. However, existing evaluations of large language models (LLMs) in wireless and networking domains largely focus on isolated tasks or treat networks as numeric constraints, leaving network-level semantic reasoning over intent, policy, trust, and multi-agent coordination insufficiently explored. This paper introduces 6G-Bench, an open benchmark for evaluating semantic communication and network-level reasoning in AI-native 6G networks. 6G-Bench defines a taxonomy of 30 decision-making tasks (T1--T30) extracted from ongoing 6G and AI-agent standardization activities in 3GPP, IETF, ETSI, ITU-T, and the O-RAN Alliance, and organizes them into five standardization-aligned capability categories. Starting from 113{,}475 scenarios, we generate a balanced pool of 10{,}000 very-hard multiple-choice questions using task-conditioned prompts that enforce multi-step quantitative reasoning under uncertainty and worst-case regret minimization over multi-turn horizons. After automated filtering and expert human validation, 3{,}722 questions are retained as a high-confidence evaluation set, while the full pool is released to support training and fine-tuning of 6G-specialized models. Using 6G-Bench, we evaluate 22 foundation models spanning dense and mixture-of-experts architectures, short- and long-context designs (up to 1M tokens), and both open-weight and proprietary systems. Across models, deterministic single-shot accuracy (pass@1) spans a wide range from 0.22 to 0.82, highlighting substantial variation in semantic reasoning capability. Leading models achieve intent and policy reasoning accuracy in the range 0.87–0.89, while selective robustness analysis on reasoning-intensive tasks shows pass@5 values ranging from 0.20 to 0.91. To support open science and reproducibility, we release the 6G-Bench dataset on GitHub: \url{https://github.com/maferrag/6G-Bench}.
\end{abstract}


% \fi
\IEEEpeerreviewmaketitle

\section{Introduction}

Emerging visions for 6G suggest that large language models (LLMs) may play a complementary role as intent-based interfaces and high-level reasoning components within AI-native network architectures, rather than as replacements for standardized network functions. This view is consistent with ongoing discussions in 3GPP, where artificial intelligence and machine learning are treated as implementation choices, while standardization efforts emphasize data exposure, analytics capabilities, and model lifecycle governance \cite{3gpp_tr_22_870_2025}. Similar perspectives are reflected in ITU-T and O-RAN Alliance studies, which identify generative AI as an enabling technology for network intelligence, intent interpretation, and cross-layer orchestration, while maintaining clear functional boundaries and interoperability requirements \cite{ITU-T-TR-GenAI-Telecom-2025,ORAN-nGRG-GenAI-6G-2025}. In this context, LLMs can be positioned above existing network functions, operating as reasoning and coordination layers that interact with standardized interfaces rather than directly controlling protocol behavior.

Beyond long-standing physical-layer advances such as massive multiple-input multiple-output (MIMO) \cite{chong2026large,yang2025large}, recent 6G research has introduced reconfigurable intelligent surfaces (RIS) \cite{huang2025llm} as a means to partially control the wireless propagation environment itself. While these technologies significantly expand the degrees of freedom available for link-level optimization, they remain fundamentally metric-driven and operate within predefined protocol abstractions. As networks evolve toward AI-native 6G architectures, the increasing programmability of radio, compute, and sensing resources exemplified by RIS-enabled environments raises decision-making challenges that extend beyond signal processing, requiring reasoning over intent, policy, trust, and cross-domain semantics.

By processing heterogeneous inputs such as network analytics, radio access network measurements, service requirements, and policy constraints, LLMs can help translate high-level intents into actionable guidance for established optimization and automation functions. This aligns with emerging work in the IETF on AI-assisted network management, intent-based interfaces, and AI agent communication protocols, which explore how reasoning agents can consume telemetry, exchange structured knowledge, and support closed-loop control without violating protocol modularity \cite{stephan,ietf_hw_ai_agent_6g_00,ietf_rosenberg_ai_protocols_00}. From a service perspective, such intent translation is particularly relevant for supporting diverse 6G service classes, including enhanced Mobile Broadband (eMBB), ultra-reliable low-latency communications (URLLC), and massive machine-type communications (mMTC), where requirements span throughput, latency, reliability, and energy efficiency across multiple domains \cite{jing2025llm}. Moreover, ongoing ETSI activities on experiential networked intelligence, multi-agent systems, and AI-native management architectures further emphasize the need for scalable, explainable, and governable AI integration in future networks \cite{etsi_gr_isc_001_v1_1_1_2025,etsi_gr_eni_051_v4_1_1_2025,etsi_eni_isg_055_early_draft_2025,etsi_gr_mat_001_v1_1_1_2026}. While current 5G-Advanced systems largely assume offline-trained models and deterministic control logic, these standardization efforts suggest that 6G may incrementally incorporate intent-based and reasoning-driven AI capabilities to enhance cross-domain coordination and service differentiation.

Recent advances toward AI-native 6G networks have highlighted the transformative role of large language models (LLMs) and foundation models in enabling semantic-aware communication, intelligent control, and autonomous network reasoning across heterogeneous environments. A growing body of work demonstrates that LLMs can move beyond traditional data-driven learning to perform high-level reasoning, decision-making, and cross-layer optimization in complex wireless systems. For example, LLM-driven frameworks have been proposed for zero-trust security automation in space–air–ground integrated networks (SAGIN) \cite{cao2025exploring}, zero-touch network security management \cite{cao2025advancing}, resilient MAC protocol design \cite{kim2025resilient}, and collaborative large–small AI architectures for green 6G operation \cite{huang2025collaborative}. In parallel, LLMs have been successfully applied to physical-layer and link-layer tasks such as predictive beamforming in near-field ISAC systems \cite{huang2025llm}, intelligent metasurface control \cite{huang2025llma}, multimodal beam prediction for V2X \cite{lei2025llm}, and semantic-level multimedia transmission \cite{jiang2025large}, illustrating their broad potential across the 6G protocol stack.

Despite these promising results, existing studies are largely evaluated in task-specific, closed experimental settings, using customized datasets, simulators, and performance metrics that hinder fair comparisons and systematic analysis. Even large-scale efforts such as SAG-Attack for SAGIN security evaluation \cite{cao2025exploring} or BATTLE-FIELD environments for automated defense validation \cite{cao2025advancing} are tailored to narrow problem domains and do not generalize across communication, sensing, security, and control tasks. Similarly, LLM-enabled solutions for V2X beamforming \cite{huang2025llm,lei2025llm}, covert NR V2X scheduling under intelligent eavesdroppers \cite{fu2025tradeoff}, and token-based MAC protocols \cite{kim2025resilient} employ disparate assumptions, input representations, and evaluation criteria. This fragmentation obscures fundamental questions about how foundation models reason over network-level semantics, how they scale across heterogeneous scenarios, and how their reasoning capability translates into measurable performance gains in AI-native 6G networks.

The key contributions of this work are summarized as follows:
\begin{itemize}
    \item We introduce 6G-Bench, an open and standardized benchmark for evaluating semantic communication and network-level reasoning with foundation models in AI-native 6G networks. 6G-Bench defines a taxonomy of 30 decision-making tasks (T1--T30) extracted from ongoing 6G and AI-agent standardization activities in 3GPP, IETF, ETSI, ITU-T, and the O-RAN Alliance, and organizes them into five capability categories: intent and policy reasoning, network slicing and resource management, trust and security awareness, AI-native networking and agentic control, and distributed intelligence and emerging 6G use cases.

    \item We design a task-conditioned MCQ construction pipeline for network-level semantic reasoning. Starting from 113{,}475 scenarios in $\alpha^3$-Bench, we develop task-specific prompts that enforce very-hard difficulty, multi-step quantitative reasoning under uncertainty, and worst-case regret minimization over multi-turn horizons. This process yields a balanced pool of 10{,}000 MCQs across all 30 tasks, generated using a heterogeneous set of state-of-the-art reasoning models and refined through automatic deduplication and anti-heuristic constraints.

    \item We establish a rigorous two-stage validation pipeline that combines automated structural and logical checks with expert human review. From the initial 10{,}000 generated questions, 3{,}722 MCQs are retained as a high-confidence evaluation set after filtering for semantic correctness, quantitative soundness, and uniqueness under worst-case reasoning. The remaining questions serve as a complementary resource for training and fine-tuning foundation models for 6G-specific semantic communication and network reasoning use cases.

    \item We curate and evaluate a diverse suite of 22 contemporary foundation models, spanning code-specialized and general-purpose systems, multimodal and long-context architectures, dense and mixture-of-experts designs, and both open-weight and proprietary models. For each model, we characterize architectural scale, context length, release timeline, and functional category, providing a deployment-oriented view of the current foundation model landscape relevant to AI-native 6G semantic networking.

    \item We propose an evaluation methodology that combines deterministic pass@1 accuracy with selective pass@k analysis on reasoning-intensive tasks, together with task- and group-level aggregation aligned with the five defined capability categories. Using this methodology, we conduct a comprehensive empirical study of all evaluated models on 6G-Bench, showing that mid-scale models can outperform larger ones in deterministic accuracy, while trust-, security-, and distributed-intelligence tasks remain the most challenging.

    \item We analyze the implications of these findings for AI-native 6G deployment and standardization, identifying which classes of foundation models are already suitable as semantic reasoning layers above standardized network functions, and where further architectural innovation, alignment strategies, or domain-specific training are required to meet the reliability, accountability, and safety expectations of future 6G networks.
\end{itemize}

The remainder of this paper is organized as follows. Section \ref{sec:related} reviews related work on benchmarking and evaluating large language models in telecommunications and wireless networking. Section \ref{sec:bench} presents the design of 6G-Bench, including the standardization-driven task taxonomy, semantic state and action abstractions, dataset construction, validation pipeline, and evaluation protocol. Section \ref{sec:perf} reports the performance evaluation of contemporary foundation models on 6G-Bench, providing task-level, group-level, and robustness (pass@k) analyses across all capability categories. Finally, Section \ref{sec:conc} concludes the paper by summarizing key findings, discussing implications for AI-native 6G deployment and standardization, and outlining directions for future research.

\begin{table*}[t]
\centering
\scriptsize
\setlength{\tabcolsep}{6pt}
\caption{Comparison of related benchmarks for standardization-aligned, network-level semantic reasoning in AI-native 6G networks.}
\label{tab:benchmark_comparison}
\begin{tabular}{lccccccc}
\hline
\textbf{Benchmark} 
& \textbf{Telecom} 
& \textbf{6G Standards} 
& \textbf{Intent / Policy} 
& \textbf{Multi-Agent} 
& \textbf{Trust \& SLA} 
& \textbf{6G Network-Aware} 
& \textbf{Semantic Reasoning} \\
\hline
TeleQnA \cite{maatouk2025teleqna}              & \checkmark & \textopenbullet & -- & -- & -- & -- & -- \\
TeleTables \cite{ezzakri2025teletables}        & \checkmark & \textopenbullet  & -- & -- & -- & -- & -- \\
TelAgentBench \cite{lee2025telagentbench}      & \checkmark & --         & \textopenbullet & \checkmark & -- & -- & --\\
LDOT \cite{lin2025go}                          & \checkmark & --         & -- & -- & -- & \textopenbullet & \textopenbullet \\
CovertComBench \cite{liu2026covertcombench}    & \checkmark & --         & -- & -- & \textopenbullet& \textopenbullet & \textopenbullet \\
\hline
\textbf{6G-Bench (Ours)}                       & \checkmark & \checkmark & \checkmark & \checkmark & \checkmark & \checkmark & \checkmark \\
\hline
\end{tabular} \\

\checkmark~indicates explicit support; 
\textopenbullet~indicates partial support; 
-- indicates not supported.

\end{table*}

\section{Related Work}
\label{sec:related}

This section reviews recent research on the systematic evaluation of large language models in the telecommunication and wireless communication domains. We organize the related work according to the primary evaluation focus of existing studies, encompassing benchmarks for general telecommunications knowledge, reasoning over structured standards artifacts such as tables and specifications, agentic capabilities in realistic telecommunication service scenarios, and advanced mathematical reasoning and constrained optimization in wireless systems. Table~\ref{tab:benchmark_comparison} presents a structured comparison between 6G-Bench and existing benchmarks, indicating whether each benchmark explicitly supports, partially supports, or does not support the key capabilities required for semantic communication and reasoning at the AI-native 6G network-level.

\subsection{Telecommunications Knowledge Benchmarks}

Maatouk et al. \cite{maatouk2025teleqna} propose TeleQnA, which is a domain-specific benchmark designed to rigorously evaluate large language models (LLMs) in telecommunications, comprising 10,000 multiple-choice questions curated from approximately 25,000 pages and 6 million words of open-access standards, research publications, surveys, and telecom lexicons. The dataset is systematically distributed across five categories, namely lexicon (5\%), research overview (20\%), research publications (45\%), standards overview (10\%), and standards specifications (20\%), in order to capture both general and highly specialized telecom knowledge. An automated question--answer generation pipeline based on two GPT-3.5 agents, acting as generator and validator, and complemented by two stages of human expert validation and clustering-based redundancy removal, ensures both scalability and quality control. Using TeleQnA, the authors benchmark GPT-3.5 and GPT-4, reporting overall accuracies of 67\% and 74\%, respectively, where GPT-4 reaches up to 87\% accuracy on lexicon-related questions but only 64\% on standards specifications, revealing persistent limitations in handling complex, standards-driven content. Moreover, the study shows that incorporating contextual information yields a 22.5\% relative accuracy gain in standards-related tasks, enabling GPT-3.5 to approach GPT-4 performance.

\subsection{Reasoning over Telecommunications Standards and Tables}

Ezzakri et al. \cite{ezzakri2025teletables} propose TeleTables, a benchmark specifically designed to assess the ability of large language models (LLMs) to interpret and reason over tabular information in telecommunications standards, with a particular focus on 3GPP technical specifications. The benchmark consists of 500 multiple-choice questions generated from 2,220 tables automatically extracted from 13 3GPP specifications spanning Releases 18 and 19, where tables are represented in multiple formats, including HTML, JSON, Markdown, and high-resolution images. A multi-stage pipeline leveraging multimodal and reasoning-oriented LLMs, followed by systematic validation and human-in-the-loop filtering, is employed to ensure correctness and controlled difficulty, resulting in a balanced set of basic and difficult questions. Extensive benchmarking across non-reasoning, multimodal, and reasoning open-weight models reveals that smaller models below 10B parameters achieve limited performance, with pass@1 scores as low as 26.15\%, while larger reasoning models such as Qwen3-32B and GPT-OSS-120B reach up to 91.18\% and 90.90\% pass@1, respectively, when optimal table representations are provided. The study further demonstrates that table format has a substantial impact on performance, with HTML consistently yielding the highest accuracy for most models, while image-only representations incur significant degradation despite high token costs. Moreover, performance decreases monotonically with increasing table complexity, as measured by HTML token length, highlighting the difficulty of reasoning over large, hierarchical tables.

\subsection{Agentic Evaluation in Telecommunication Service Scenarios}

Lee et al. \cite{lee2025telagentbench} propose TelAgentBench, a multifaceted industry-oriented benchmark to systematically evaluate agents of large language models (LLM) in realistic telecommunication service environments. The benchmark comprises more than 1{,}700 synthetically generated and expert-validated instances in Korean, designed to assess five core agentic capabilities: Reasoning, Planning, Action (tool use), Retrieval-Augmented Generation (RAG), and Instruction Following. Each capability is instantiated through a dedicated dataset module, including 225 multi-hop reasoning questions, 200 multi-constraint planning tasks, 757 tool-calling scenarios spanning up to 23 business support system APIs, 258 RAG instances with realistic distractor documents, and 300 multi-turn instruction-following dialogues incorporating telecom-specific and linguistic constraints. The dataset construction follows a four-stage pipeline encompassing scenario design from real-world telecom use cases, development of domain-specific execution and retrieval environments, large-scale data generation and augmentation, and rigorous human-in-the-loop validation by telecom service agents and language experts. Extensive evaluation across 15 proprietary and open-source LLMs reveals consistent and significant performance gaps between thinking-enabled and non-thinking models, with explicit reasoning improving accuracy across all dimensions, for example, exceeding 85\% accuracy in reasoning tasks and yielding up to a 12\% gain in action-oriented evaluations.

\subsection{Reasoning and Optimization in Wireless Communications}

Lin et al. \cite{lin2025go} propose the LDOT benchmark for evaluating the advanced reasoning and domain-specific capabilities of large language models (LLMs) in wireless communications, beyond the limits of existing telecom datasets. LDOT comprises 2{,}341 carefully filtered question--answer pairs spanning three categories: 208 conceptual questions, 1{,}650 mathematical and wireless logical problems, and 483 complex network optimization tasks, with optimization questions exhibiting an average length of 2{,}573.9 characters and requiring detailed multi-step derivations. The dataset is constructed through a multi-stage pipeline that combines large-scale literature retrieval, data sanitization, automated multi-agent question generation, agentic retrieval-augmented generation for optimization problems, adversarial filtering, and expert human validation. Extensive evaluations of a wide range of open-source and closed-source LLMs, including GPT-4o, DeepSeek R1, Qwen2.5-32B, and o3-mini, reveal that while general-purpose models achieve reasonable performance on conceptual telecom questions, they perform poorly on reasoning-intensive mathematical and optimization tasks, with the best-performing model achieving only 0.440 accuracy under strict evaluation criteria. The analysis further demonstrates that retrieval augmentation substantially improves performance on conceptual questions but offers limited gains for optimization problems, where failures stem primarily from inadequate reasoning rather than missing knowledge.

\subsection{Security-Constrained Wireless Communication Benchmarks}

Liu et al. \cite{liu2026covertcombench} introduce \emph{CovertComBench}, a domain-specific benchmark designed to rigorously evaluate large language models in wireless covert communication (CC), where performance optimization must satisfy strict detection-theoretic constraints. Unlike conventional communication tasks, CC requires maximizing transmission utility under a covertness constraint formulated via Kullback--Leibler divergence bounds, posing a challenging constrained optimization problem. CovertComBench is a fully human-verified benchmark comprising 517 questions across three task categories: Multiple-Choice Questions (MCQs), Optimization Derivation Questions (ODQs), and Code Generation Questions (CGQs), covering modern CC scenarios including IRS-, NOMA-, and MIMO-based systems. Extensive evaluation in state-of-the-art LLMs shows strong performance in conceptual understanding (up to 81\% accuracy) and code generation (up to 83\%), but substantially weaker results in mathematical derivation tasks, with ODQ accuracy ranging from 18\% to 55\%. The study further analyzes the reliability of automated evaluation using an LLM-as-Judge framework, revealing notable discrepancies from expert scoring in mathematically intensive tasks. These findings indicate that current LLMs are effective implementation assistants, but remain unreliable as autonomous solvers for security-constrained wireless optimization, motivating future integration with external symbolic computation tools.

\subsection{Positioning of 6G-Bench}

The benchmarks reviewed above have substantially advanced the evaluation of large language models in telecommunications, yet they remain limited in scope with respect to AI-native 6G requirements. Knowledge-centric benchmarks such as TeleQnA  \cite{maatouk2025teleqna} primarily assess factual understanding of telecom concepts and standards, without evaluating decision-making under evolving network conditions. Table- and specification-oriented benchmarks like TeleTables \cite{ezzakri2025teletables} focus on structured artifact interpretation, but abstract away the semantic dynamics of intent, policy, and network control. Agentic benchmarks such as TelAgentBench \cite{lee2025telagentbench} evaluate service-level workflows and tool usage, but do not explicitly model network-level semantics, worst-case reasoning, or multi-turn decision regret under uncertainty. Similarly, optimization- and security-focused benchmarks (e.g., LDOT  \cite{lin2025go}  and CovertComBench \cite{liu2026covertcombench}) target mathematically constrained problems or narrowly defined communication scenarios, rather than holistic network-level reasoning across heterogeneous 6G operational domains.

In contrast, \emph{6G-Bench} is explicitly designed to evaluate semantic communication and network-level reasoning as first-class capabilities in AI-native 6G networks. The benchmark departs from metric-driven or task-isolated evaluations by grounding all benchmark tasks in meaning-bearing abstractions such as intents, policies, slices, trust relationships, and agent coordination. Its taxonomy of 30 tasks is systematically extracted from ongoing 6G and AI-agent standardization activities in 3GPP, IETF, ETSI, ITU-T, and the O-RAN Alliance, ensuring that benchmarked capabilities directly reflect emerging architectural and operational requirements rather than ad hoc problem formulations.

Moreover, 6G-Bench uniquely emphasizes worst-case, uncertainty-aware, and multi-turn reasoning through task-conditioned multiple-choice questions derived from realistic episodes, with oracle decisions defined via regret minimization rather than instantaneous performance optimization. This design enables direct assessment of whether foundation models can support deployment-relevant decisions in safety-, trust-, and SLA-critical contexts. As such, 6G-Bench complements existing telecom benchmarks by filling a critical gap: the systematic, standardized evaluation of foundation models as semantic reasoning layers above network functions in AI-native 6G systems, rather than as isolated solvers or knowledge engines.

\begin{figure*}[t]
    \centering
    \includegraphics[width=1\textwidth]{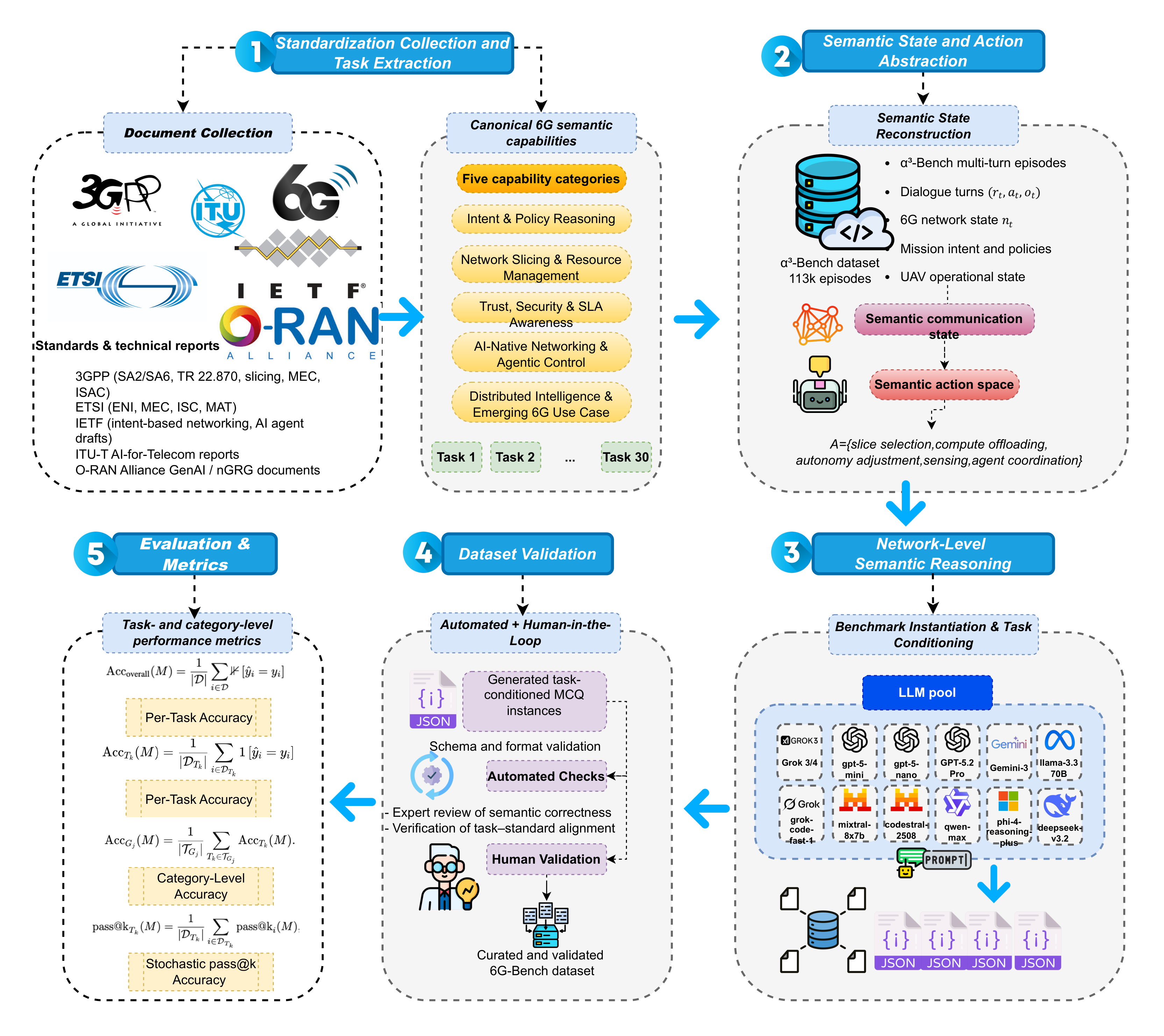}
    \caption{Overview of the 6G-Bench benchmark design, task construction, validation, and evaluation pipeline.}
    \label{fig:6gbench_pipeline}
\end{figure*}

\begin{algorithm}[t]
\caption{6G-Bench Benchmark Construction}
\label{alg:6gbench_construction_compact}
\DontPrintSemicolon
\KwIn{
Standards $\mathcal{S}$; episode set $\mathcal{E}$ from $\alpha^3$-Bench / UAVBench
}
\KwOut{
Benchmark dataset $\mathcal{D}$
}

Extract semantic capabilities from $\mathcal{S}$ and define tasks
$\{T_1,\dots,T_{30}\}$ grouped into $\mathcal{G}=\{G_1,\dots,G_5\}$.\;
$\mathcal{D} \leftarrow \emptyset$\;

\ForEach{episode $e \in \mathcal{E}$}{
    Parse dialogue turns $d_t=(r_t,a_t,o_t,n_t)$.\;
    Reconstruct semantic states $s_t=(i_t,n_t,p_t,x_t)$.\;

    Identify decision turns $\mathcal{T}_e$.\;
    \ForEach{$t \in \mathcal{T}_e$}{
        Determine action space $\mathcal{A}_t$.\;
        Obtain oracle action $a_t^\star$ via worst-case regret minimization.\;
        Form truncated trajectory $s_{1:t}$.\;
        Construct MCQ with candidates $\mathcal{A}_t=\{A,B,C,D\}$ and label $y$.\;
        Assign task identifier $T_k$.\;
        Validate instance (automatic + expert review).\;
        \If{valid}{
            Add $(S_e,q,y,T_k)$ to $\mathcal{D}$.\;
        }
    }
}
\Return $\mathcal{D}$\;
\end{algorithm}

\begin{figure*}[t]
  \centering
  \begin{subfigure}{0.32\textwidth}
    \centering
     \includegraphics[width=\linewidth]{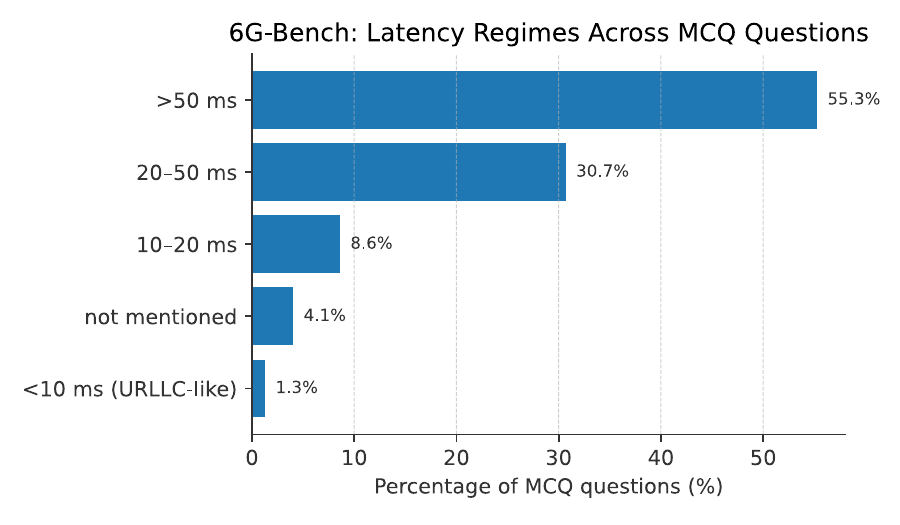}
    \caption{Latency regimes.}
    \label{fig:latency-regimes}
  \end{subfigure}
  \hfill
  \begin{subfigure}{0.32\textwidth}
    \centering
     \includegraphics[width=\linewidth]{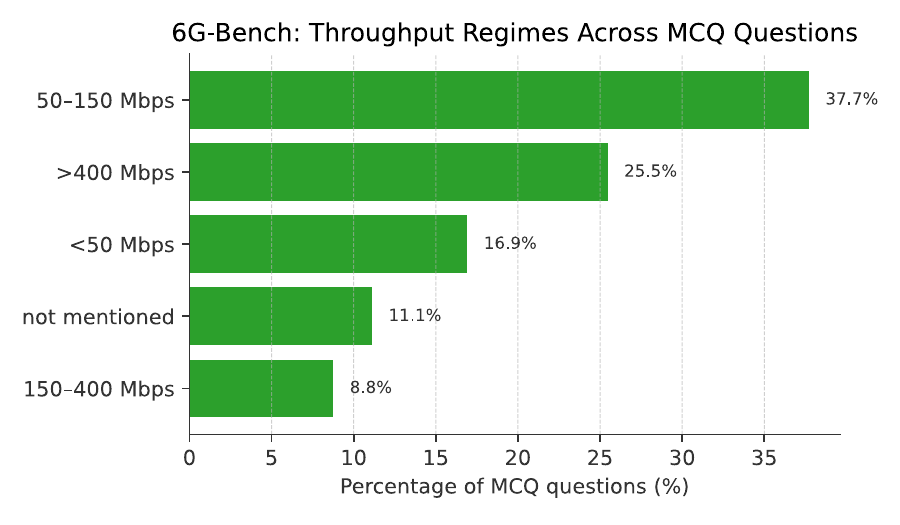}
    \caption{Throughput regimes.}
    \label{fig:throughput-regimes}
  \end{subfigure}
  \hfill
  \begin{subfigure}{0.32\textwidth}
    \centering
     \includegraphics[width=\linewidth]{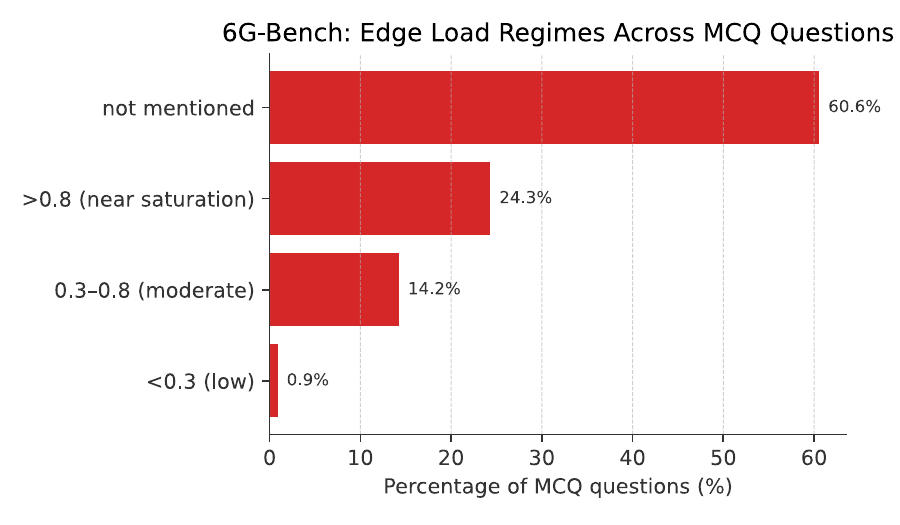}
    \caption{Edge-compute load regimes.}
    \label{fig:edge-load-regimes}
  \end{subfigure}

  \caption{Network condition diversity in 6G-Bench.
  Subfigures (a)--(c) show the proportion of MCQ questions operating under
  different latency, throughput, and edge-compute load regimes.}
  \label{fig:6gbench-network-diversity}
\end{figure*}

\section{Benchmark Design and Evaluation Methodology}
\label{sec:bench}

Figure~\ref{fig:6gbench_pipeline} presents an overview of the 6G-Bench benchmark design and evaluation pipeline.
The process begins by collecting 6G-related standardization and technical reports from 3GPP, ETSI, IETF, ITU-T, and the O-RAN Alliance, from which canonical semantic capabilities are extracted and organized into five capability categories and a task taxonomy of 30 tasks.
Using multi-turn episodes from the $\alpha^3$-Bench dataset, 6G-Bench reconstructs semantic communication states and corresponding semantic action spaces that expose intent-, policy-, and network-level abstractions.
The Network-level semantic reasoning is then applied to instantiate multiple-choice benchmark instances conditioned on the task.
Prior to evaluation, all generated instances undergo automated validation and human-in-the-loop expert review to ensure semantic correctness and alignment with standardization intent.
Finally, models are evaluated using task-level and category-level performance metrics, enabling fine-grained assessment of semantic reasoning capabilities across heterogeneous 6G operational domains.

\subsection{Problem Formulation and Episode Source}

The Algorithm~\ref{alg:6gbench_construction_compact} summarizes the end-to-end construction of 6G-Bench, from standardization-driven capability extraction and semantic state reconstruction to task-conditioned multiple-choice benchmark instantiation and validation.

\subsubsection{Episode Source}

6G-Bench is constructed on top of $\alpha^3$-Bench \cite{ferrag2026alpha} and UAVBench\cite{ferrag2025uavbench}, a large-scale benchmark that models autonomous UAV missions as multi-turn, language-mediated control loops operating under dynamic 6G network conditions.
Each episode corresponds to a structured conversational interaction between an LLM-based UAV agent and a supervisory entity, where decisions must satisfy mission objectives, safety policies, protocol constraints, and time-varying network conditions.

Formally, an episode is defined as a finite dialog trajectory:
\begin{equation}
\mathcal{E} = \{ d_t \}_{t=1}^{T},
\end{equation}
where each dialog turn $d_t$ is represented as:
\begin{equation}
d_t = (r_t, a_t, o_t, n_t).
\end{equation}

See \cite{ferrag2026alpha} for details on episode generation, validation, and network modeling. 

\subsubsection{Semantic Network State}

At each dialog turn $t$, the 6G network context is represented by the semantic network state vector:
\begin{equation}
n_t = (\text{slice}_t,\; \ell_t,\; j_t,\; \rho_t,\; \tau_t,\; e_t),
\end{equation}
where $\text{slice}_t \in \{\text{URLLC}, \text{eMBB}, \text{mMTC}\}$ denotes the active network slice, $\ell_t$ is the end-to-end latency, $j_t$ the jitter, $\rho_t$ the packet loss rate, $\tau_t$ the achievable throughput, and $e_t \in [0,1]$ the normalized edge-compute load. These parameters are explicitly injected into the conversational context and evolve dynamically during the episode.

Figure~\ref{fig:6gbench-network-diversity} characterizes the operating regimes under which benchmark decisions are instantiated, providing a regime-level view of network stress conditions. As shown in Figure~\ref{fig:latency-regimes}, severely degraded latency exceeding 50\,ms appears in 55.3\% of the questions, while an additional 30.7\% operate in the intermediate 20–50\,ms range, indicating that the benchmark predominantly targets performance-critical and degraded conditions rather than nominal operation. Figure~\ref{fig:throughput-regimes} illustrates that achievable throughput spans from constrained regimes below 50\,Mbps (16.9\%) to high-capacity regimes above 400\,Mbps (25.5\%), reflecting realistic transitions between URLLC-dominated safety phases and eMBB-oriented data-intensive operation. Figure~\ref{fig:edge-load-regimes} further shows that near-saturated edge-compute load conditions above 0.8 occur in 24.3\% of the instances, while moderate load regimes remain common, underscoring that many tasks—such as graceful degradation, preemptive autonomy downgrade, and compute placement—are evaluated under edge contention consistent with ETSI MEC and AI-native orchestration assumptions \cite{etsi_gs_mec_003_v3_1_1_2022}.

\subsubsection{Semantic Communication Model}

We define semantic communication as the exchange of meaning-bearing abstractions rather than raw symbols.
Accordingly, the semantic state at turn $t$ is defined as:
\begin{equation}
s_t = (i_t,\; n_t,\; p_t,\; x_t),
\end{equation}
where $i_t$ represents mission or network intent, $n_t$ is the 6G network state, $p_t$ denotes policy and service-level constraints, and $x_t$ captures the UAV operational state (e.g., battery level, speed, and position). This state captures the information communicated at the intent, policy, and network levels in AI-native 6G systems.

\begin{figure*}[t]
  \centering
   \includegraphics[width=\linewidth]{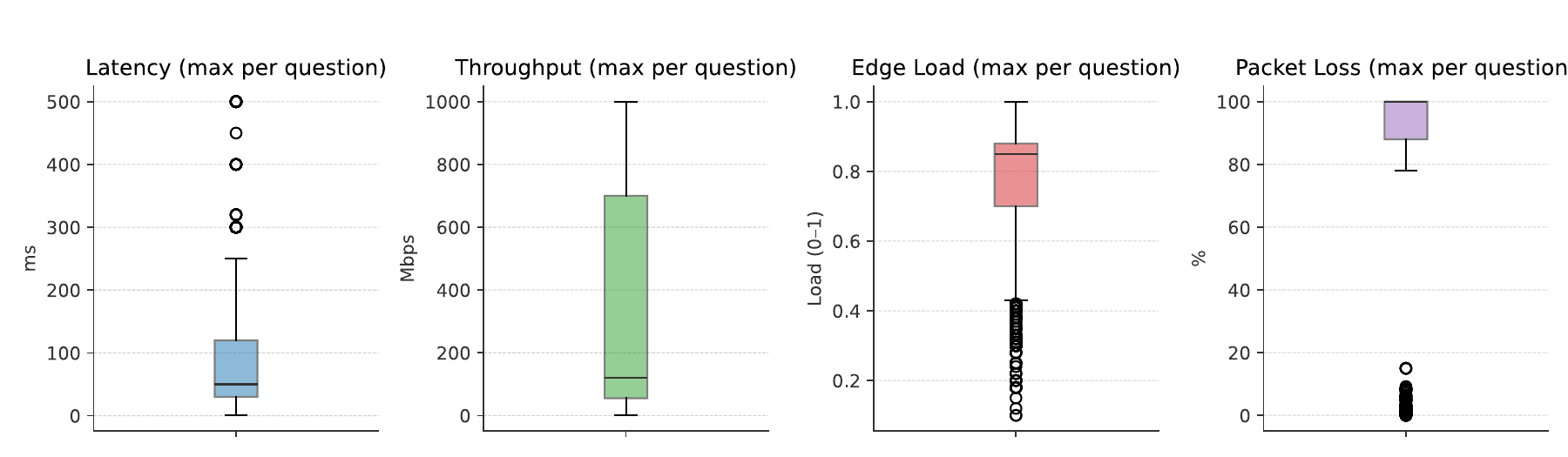}
  \caption{Distribution of per-question maximum network metrics in 6G-Bench,
  including latency, throughput, edge-compute load, and packet loss.}
  \label{fig:6gbench-max-network-metrics}
\end{figure*}

Figure~\ref{fig:6gbench-max-network-metrics} complements the regime-level analysis by presenting the distribution of per-question maximum network metrics, thereby exposing the extremal conditions under which oracle decisions are derived. The boxplots reveal heavy-tailed, highly dispersed distributions for latency, throughput, edge-compute load, and packet loss, with maximum latency exceeding several hundred milliseconds and edge-compute load frequently approaching full saturation. Packet loss similarly reaches extreme values, confirming that a substantial subset of benchmark instances requires reasoning under severe reliability degradation. These observations demonstrate that 6G-Bench systematically emphasizes worst-case and uncertainty-aware decision-making, directly aligning the dataset construction with the minimax regret objective used to determine oracle actions and reflecting deployment-relevant conditions in AI-native 6G networks, where future network evolution, policy compliance, and safety margins must be anticipated rather than reactively optimized.

\subsubsection{Decision Space and State Evolution}

At each turn, the agent selects a semantic action:
\begin{equation}
a_t \in \mathcal{A},
\end{equation}
where $\mathcal{A}$ includes slice selection, compute offloading, autonomy adjustment, sensing requests, and agent-to-agent coordination.

The system evolves according to:
\begin{equation}
s_{t+1} = f(s_t, a_t, \xi_t),
\end{equation}
where $f(\cdot)$ denotes the environment transition function and $\xi_t$ captures uncertainty arising from network variability, sensing noise, and environmental disturbances.

\subsubsection{Objective: Network-Level Semantic Reasoning}

Rather than optimizing instantaneous performance, agents must reason over future consequences under uncertainty.
Given a finite horizon $H$, the decision objective is formulated as:
\begin{equation}
a_t^\star = \arg\min_{a \in \mathcal{A}} \max_{\xi_{t:t+H}} \sum_{k=t}^{t+H} C(s_k, a_k),
\end{equation}
where $C(\cdot)$ captures mission degradation, safety risk, and SLA violation penalties.

This formulation reflects network-level semantic reasoning, where decisions are evaluated based on worst-case future regret rather than immediate network metrics.

\subsubsection{Benchmark Instantiation in 6G-Bench}

6G-Bench instantiates this formulation by transforming $\alpha^3$-Bench episodes into task-conditioned multiple-choice reasoning problems.
Each benchmark instance corresponds to:
\begin{equation}
(s_{1:t},\; \mathcal{A}_t,\; a_t^\star),
\end{equation}
where $s_{1:t}$ is a truncated semantic trajectory, $\mathcal{A}_t = \{a^{(A)}, a^{(B)}, a^{(C)}, a^{(D)}\}$ is a finite set of candidate actions, and $a_t^\star$ minimizes worst-case regret.

A model succeeds if it correctly identifies $a_t^\star$ based solely on semantic communication artifacts and network-aware reasoning.

% In preamble:
% \usepackage{graphicx}
% \usepackage{subcaption}

\subsection{Task Taxonomy and Standardization-Aligned Categories}

To evaluate semantic communication and network-level reasoning in AI-native 6G systems, 6G-Bench defines a set of 30 tasks (T1--T30) organized into five capability-oriented categories.
These categories reflect recurring architectural themes and control challenges identified in ongoing 6G and AI agent standardization activities in 3GPP \cite{3gpp_tr_22_870_2025}, IETF \cite{stephan,ietf_hw_ai_agent_6g_00,ietf_rosenberg_ai_protocols_00}, ETSI\cite{etsi_gr_isc_001_v1_1_1_2025,etsi_gr_eni_051_v4_1_1_2025,etsi_eni_isg_055_early_draft_2025,etsi_gr_mat_001_v1_1_1_2026}, ITU-T \cite{ITU-T-TR-GenAI-Telecom-2025} and the O-RAN Alliance \cite{ORAN-nGRG-GenAI-6G-2025}.
Rather than targeting isolated network functions, the taxonomy emphasizes decision-making over meaning-bearing abstractions such as intents, policies, slices, trust relationships, and agent coordination.

\subsubsection{Intent \& Policy Reasoning Performance}

Intent interpretation and policy alignment form the foundation of AI-native network operation \cite{ETSI_TS_128_312_v17_0_1}. In the intent-based networking frameworks studied by the IETF and 3GPP SA2, high-level intents express desired mission or service outcomes, while the network remains responsible for assessing feasibility, enforcing policies, and adapting execution as conditions evolve. As a result, intents must be continuously evaluated against network state, operational constraints, and policy rules throughout the mission lifecycle \cite{ETSI_TS_128_105_v17_3_0}. In realistic deployments, network observability is often imperfect due to missing signals, delayed KPIs, or noisy sensing streams. Consequently, intent- and policy-aware reasoning must explicitly account for telemetry uncertainty rather than assuming complete or timely network state information.

From a semantic communication perspective, intent represents a persistent abstraction that bears meaning and spans multiple dialog turns. An AI-native agent \cite{xiao2026towards} must not only interpret the initial intent correctly, but also ensure that subsequent decisions remain consistent with that intent as network conditions, uncertainty, and policy constraints change \cite{ETSI_TS_128_104_v17_0_1}. Failures in this capability may lead to infeasible mission execution, implicit policy violations, or inconsistent decisions between replanning events \cite{ETSI_TS_128_313_v16_0_0}. 

The following tasks explicitly evaluate these aspects of intent- and policy-aware reasoning \cite{jiang2026agentic}:
\begin{itemize}
    \item \textbf{T1: Intent Feasibility Assessment} evaluates whether a model can determine if a given mission and associated 6G intent message are feasible under current network conditions, and whether adjustments are required to maintain compliance.
    \item \textbf{T2: Intent Conflict Resolution} examines the ability to resolve conflicts between mission intent and network or operator policies, requiring semantic reconciliation rather than simple rejection.
    \item \textbf{T3: Intent Drift Detection} assesses whether a model can detect subtle changes in network intent that emerge during mission execution, even when no explicit intent update is issued.
    \item \textbf{T12: Conservative Continuation Decision} evaluates whether the model can decide to continue, hold, or abort a mission when network state or sensing information is incomplete, delayed, or uncertain, balancing safety and intent preservation against premature termination or unsafe persistence.
    \item \textbf{T15: Decision Consistency under Replanning} measures the ability to avoid contradictory decisions across turns when replanning occurs due to updated intent interpretation, network feedback, or policy constraints.
\end{itemize}

\begin{figure*}[t]
  \centering
  % (a) Questions per capability category
  \begin{subfigure}{0.48\textwidth}
    \centering
     \includegraphics[width=\linewidth]{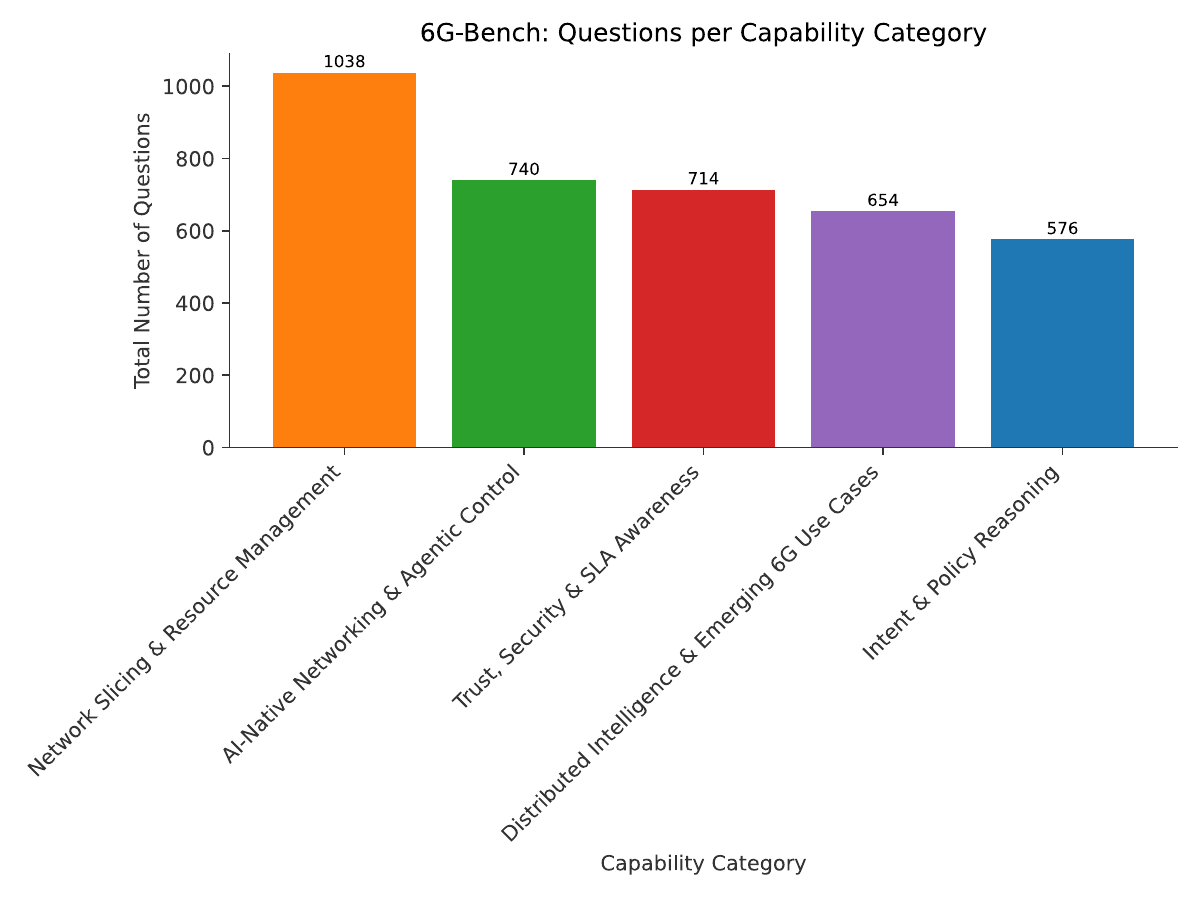}
    \caption{Questions per capability category.}
    \label{fig:tasks-per-category}
  \end{subfigure}
  \hfill
  % (b) Questions per task
  \begin{subfigure}{0.48\textwidth}
    \centering
     \includegraphics[width=\linewidth]{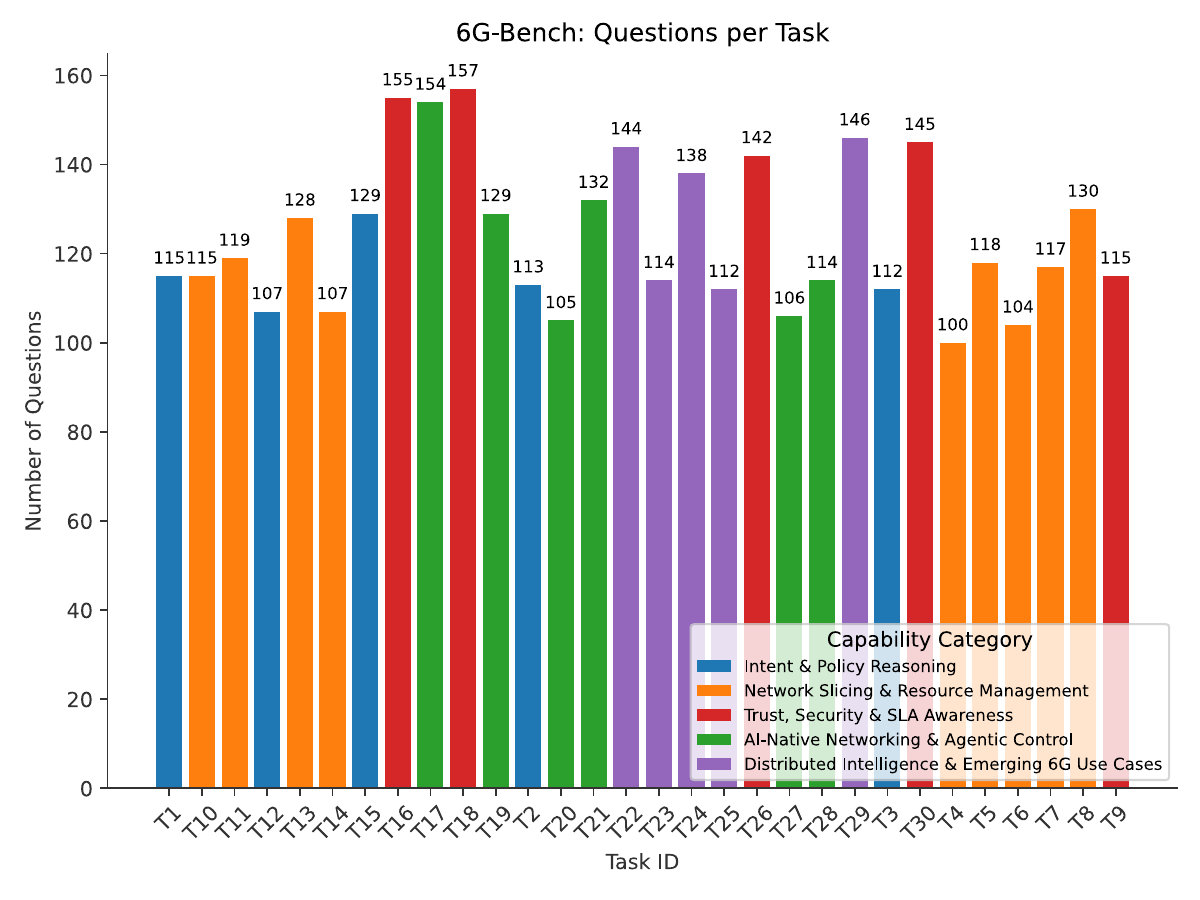}
    \caption{Questions per task (T1--T30).}
    \label{fig:tasks-per-task}
  \end{subfigure}

  \caption{Coverage of 6G-Bench across tasks and capability categories.
  Subfigure~\subref{fig:tasks-per-category} shows that questions are
  distributed across the five standardization-aligned capability categories, while
  Subfigure~\subref{fig:tasks-per-task} details the number of MCQ
  instances for each individual task T1--T30.}
  \label{fig:6gbench-task-coverage}
\end{figure*}

\subsubsection{Network Slicing \& Resource Management Performance}

Adaptive allocation of communication and compute resources lies at the core of AI-native 6G system design \cite{zou2026large}.
Through service differentiation across ultra-reliable low-latency communication (URLLC), enhanced mobile broadband (eMBB), and massive machine-type communication (mMTC), 3GPP architectures enable heterogeneous performance guarantees tailored to mission and service semantics.
At the same time, the ETSI MEC frameworks \cite{etsi_gs_mec_003_v3_1_1_2022} highlight the importance of flexible computing placement and proactive workload adaptation to sustain service levels under edge congestion and fluctuating demand.

In this context, semantic reasoning about resources requires more than reacting to instantaneous metrics.
An AI-native agent must interpret network signals in light of mission objectives, multi-agent interactions, and anticipated future conditions, often choosing stability or controlled degradation over short-term optimization  \cite{3gpp_tr_22_870_2025,etsi_gr_isc_001_v1_1_1_2025,etsi_gr_mat_001_v1_1_1_2026,ORAN-nGRG-GenAI-6G-2025}. Energy efficiency and environmental sustainability have emerged as first-class objectives in 6G operator strategies. AI-native agents must therefore reason not only about latency, throughput, and SLA compliance, but also about energy consumption and carbon impact when selecting slices, placing compute workloads, or scheduling inference and training across device, edge, and cloud resources. Resource-aware reasoning is examined through a set of tasks that capture distinct but interrelated decision challenges:
\begin{itemize}
    \item \textbf{T4: Slice Selection Reasoning} probes whether a model can choose among URLLC, eMBB, or hybrid slicing options and justify the selection based on mission requirements and network characteristics \cite{chowdhury2025framework}.
    \item \textbf{T5: Slice Switching Decision} focuses on the ability to recognize scenarios in which refraining from slice switching is preferable, even when degradation is observed.
    \item \textbf{T6: Slice Fairness vs.\ Safety} examines how contention among multiple agents is resolved when fairness objectives conflict with safety-critical needs.
    \item \textbf{T7: Compute Placement Decision} evaluates reasoning over onboard versus edge- or cloud-based inference under latency, reliability, SLA, and energy efficiency constraints, including trade-offs between performance and energy or carbon-aware operation.
    \item \textbf{T8: Graceful Degradation under Edge Overload} assesses whether autonomy or service quality can be reduced specifically in response to imminent edge overload or SLA violation risk, with the goal of stabilizing compute or network resources while preserving mission continuity.
    \item \textbf{T10: SLA Violation Prediction} measures the ability to forecast impending SLA violations from early network signals rather than relying on reactive responses.
    \item \textbf{T11: Preemptive Autonomy Downgrade} examines proactive reductions in autonomy triggered by anticipated failures beyond pure resource overload, including security risks, mobility instability, battery depletion, or channel quality deterioration.
    \item \textbf{T13: Swarm-Level Slice Negotiation} captures collaborative reasoning over slice allocation among multiple agents sharing constrained network resources.
    \item \textbf{T14: Scheduler Reconfiguration Adaptation} evaluates the consistency of decisions following changes in network or AI scheduling behavior.
\end{itemize}

\subsubsection{Trust, Security \& SLA Awareness Performance}

As AI-native networks increasingly expose control interfaces, data streams, and compute resources to autonomous agents, performance can no longer be considered in isolation from trust and security.
Current directions in IETF \cite{ietf_rosenberg_ai_protocols_00}, ETSI\cite{etsi_gr_eni_051_v4_1_1_2025,etsi_eni_isg_055_early_draft_2025}, and ITU-T \cite{ITU-T-TR-GenAI-Telecom-2025} emphasize zero-trust networking, policy-driven authorization, and service-level agreement (SLA) enforcement as foundational mechanisms for safe and accountable network operation. 

Within this setting, semantic reasoning must account for \emph{who} is requesting access, \emph{what} is being exposed, and \emph{under which guarantees} services are delivered. Decisions that optimize latency or throughput may still be unacceptable if they violate trust assumptions, regulatory constraints, or contractual SLAs \cite{mahmood2025securing,zhang2025explainable}. In addition to adversarial threats, AI-native networks must handle non-malicious faults such as hardware failures, software crashes, misconfigurations, and cascading outages, which require distinct reasoning and recovery strategies.

Trust- and security-aware reasoning is explored through tasks that reflect these trade-offs:
\begin{itemize}
    \item \textbf{T9: Trust-Aware Offloading} evaluates whether a model can correctly reject edge offloading requests when trust, policy, or authorization constraints outweigh performance benefits.
    \item \textbf{T16: Network-Exposed Compute Marketplace} examines reasoning about whether and how operator edge or cloud compute resources should be exposed to third parties, including pricing and allocation considerations.
    \item \textbf{T18: AI Agent Identity \& Onboarding} assesses the ability to authorize, authenticate, and register AI agents, as well as to decide appropriate identity mapping and digital representation policies.
    \item \textbf{T26: Trust-Aware Third-Party Agent Exposure} evaluates decisions on the level of data, API, or compute exposure granted to third-party agents based on trust relationships, regulatory constraints (e.g., data residency, lawful intercept), operator policies, and user consent.
    \item \textbf{T30: Network Security Detection \& Response Automation} examines whether a model can reason about automated detection, isolation, and recovery actions for both adversarial incidents and non-adversarial network faults, ensuring service continuity while respecting policy and SLA constraints.
\end{itemize}

\subsubsection{AI-Native Networking \& Agentic Control Performance}

Emerging 6G architectures increasingly position AI agents as integral components of the network control plane rather than as external applications.
This shift, reflected in O-RAN \cite{ORAN-nGRG-GenAI-6G-2025}, 3GPP SA6, and ITU discussions, envisions AI-driven orchestration, closed-loop automation, and agent-based interaction as core mechanisms for network operation and service management.

Within such agent-centric environments, semantic reasoning extends beyond local decision-making to include coordination, knowledge exchange, and lifecycle-aware control across multiple agents and network domains.
An AI-native system must determine not only how agents act, but also how they communicate, collaborate, and evolve over time under policy and resource constraints.

These capabilities are examined through tasks that capture key aspects of agentic control \cite{stephan}:
\begin{itemize}
    \item \textbf{T17: Network-Knowledge RAG Augmentation} evaluates whether a model can decide which network telemetry and knowledge should be exposed to support retrieval-augmented generation \cite{chen2025toward,salan2025rag} under latency and policy constraints.
    \item \textbf{T19: AI Agent Interoperability \& Federation} examines reasoning about compatibility, data sharing, and coordination when multiple AI agents from different administrative domains or operators must collaborate, including inter-operator trust models, roaming constraints, and cross-domain policy alignment.
    
    \item \textbf{T20: Agent-to-Agent Communication Management} assesses decisions on routing, quality of service, and security policies for horizontal communication between AI agents \cite{a2a_protocol_latest}, either directly or via network relays.
    \item \textbf{T21: Device-Network Task Offload Arbitration} evaluates whether AI tasks should be executed on-device, offloaded to peer agents, or delegated to edge or cloud resources, given latency, energy, capability, and trust considerations.
    \item \textbf{T27: Agent Lifecycle \& Management} examines reasoning about agent instantiation, scaling, migration, and retirement in accordance with operator policies and SLA constraints.
    \item \textbf{T28: 6G Model Training-as-a-Service Decision} evaluates whether network-facilitated model training requests should be accepted, considering resource availability, privacy implications, and quality-of-service impact.
\end{itemize}

\subsubsection{Distributed Intelligence \& Emerging 6G Use Case Performance}

Beyond conventional data transport, 6G networks are expected to enable distributed intelligence, integrated sensing, and mission-critical services.
Current 3GPP \cite{3gpp_tr_22_870_2025},
ETSI \cite{etsi_gr_isc_001_v1_1_1_2025,etsi_gr_mat_001_v1_1_1_2026},
ITU-T \cite{ITU-T-TR-GenAI-Telecom-2025}, and
O-RAN Alliance \cite{ORAN-nGRG-GenAI-6G-2025} studies highlight use cases such as integrated sensing and communication (ISAC) \cite{hong2026integrated}, digital twins assisted by the network \cite{naeem2025survey}, and public safety operations, where communication, perception, and reasoning are closely intertwined.

In these scenarios, semantic reasoning must span multiple agents, heterogeneous data sources, and time-sensitive network services.
Decisions often involve coordinating learning, sensing, and control across distributed entities while respecting latency, privacy, and resource constraints.

Such capabilities are evaluated through tasks that reflect emerging 6G application domains:
\begin{itemize}
    \item \textbf{T22: Federated / Collaborative Learning Orchestration} examines whether a model can decide when and how to schedule federated training or collaborative model updates across devices and edge nodes while meeting privacy and resource limits \cite{jain2026tinyfed6g}.
    \item \textbf{T23: Network-Assisted Digital Twin Control} evaluates reasoning about how the network should supply sensing, telemetry, and command channels to maintain and act on real-time digital twins \cite{ferrag2023poisoning}.
    \item \textbf{T24: Sensing-Enhanced Decisioning (ISAC)} assesses decisions on which sensing streams and fusion strategies the network should provide to support time-sensitive perception tasks \cite{zhu2024enabling}.
    \item \textbf{T25: AI-Agent-based Disaster / Public-Safety Coordination} examines coordination among agents, slice and sensing allocation, and command escalation policies in disaster response and public-safety scenarios.
    \item \textbf{T29: Immersive / AR Resource Prioritization} evaluates allocation of network slices and edge resources to multi-modal immersive sessions, balancing throughput, latency, and fairness \cite{su2026resource}.
\end{itemize}

Figure~\ref{fig:6gbench-task-coverage} summarizes the structural coverage of 6G-Bench across its task taxonomy and capability-oriented categories. Figure~\ref{fig:tasks-per-category} shows the distribution of the 3,722 MCQ instances across the five standardization-aligned capability groups, confirming that the benchmark spans intent and policy reasoning, network slicing and resource management, trust and security awareness, AI-native agentic control, and distributed intelligence for emerging 6G use cases. Figure~\ref{fig:tasks-per-task} further details the number of instances assigned to each of the 30 benchmark tasks (T1–T30), demonstrating that no task is sparsely represented. This distribution ensures that evaluation results reflect a broad spectrum of semantic decision-making challenges, including intent feasibility assessment, slice selection and negotiation, trust-aware offloading, agent lifecycle management, and network-assisted digital twin control, rather than being driven by a narrow subset of network functions.

\begin{algorithm}[t]
\caption{6G-Bench Evaluation Protocol}
\label{alg:6gbench_evaluation_compact}
\DontPrintSemicolon
\KwIn{
Dataset $\mathcal{D}$; model $M$; optional $k$ for pass@k
}
\KwOut{
Accuracy and pass@k metrics
}

Initialize counters for overall, per-task, and group-level statistics.\;

\ForEach{instance $(S_e,q,y,T_k) \in \mathcal{D}$}{
    Build standardized task-conditioned prompt.\;
    Query model $M$ with deterministic decoding.\;
    Parse prediction $\hat{y} \in \{\text{A},\text{B},\text{C},\text{D}\}$ using robust extraction.\;
    Update overall and per-task correctness counts.\;
    Store evaluation record for analysis.\;
}

Compute overall accuracy $\mathrm{Acc}_{\text{overall}}$.\;
Compute per-task accuracies $\mathrm{Acc}_{T_k}$.\;
Compute group-level accuracies as unweighted means over tasks in each $G_j$.\;

\If{$k>1$}{
    \ForEach{instance in reasoning-intensive subset}{
        Sample $k$ stochastic predictions.\;
        Mark pass@k if any sample matches ground truth.\;
    }
    Compute pass@k statistics overall and per task.\;
}

\Return accuracy metrics and evaluation logs.\;
\end{algorithm}

\subsection{Evaluation Protocol and Metrics}

6G-Bench is designed to evaluate models in a unified, task-aware, and episode-conditioned setting.
Each evaluation instance is a multiple-choice question (MCQ) derived from an $\alpha^3$-Bench episode and annotated with a target task $T_k$ from the taxonomy described above. Algorithm~\ref{alg:6gbench_evaluation_compact} specifies the unified evaluation protocol of 6G-Bench, including task-conditioned prompt construction, robust answer extraction, and the computation of overall, per-task, group-level, and pass@k performance metrics.

\subsubsection{Episode--Question Pairs}

The dataset is organized as paired files containing raw episodes and their corresponding MCQs.
For each episode $e$, we store:
\begin{itemize}
    \item the original dialogue and initial state (environment, airspace, UAV, policy), and
    \item a set of MCQs $\{q_{e,1}, \dots, q_{e,n_e}\}$, each associated with a task identifier $t \in \{T1,\dots,T30\}$ and a correct answer $y \in \{\text{A},\text{B},\text{C},\text{D}\}$.
\end{itemize}

During evaluation, we only consider episodes for which both the raw episode file and the MCQ file are present.
For each such episode, a compact textual summary is constructed using the same procedure as in the MCQ generation stage: initial state fields (environment, airspace, UAV, policy, success) are followed by a truncated dialogue trace (up to 12 turns), including intents, actions, observations, and network telemetry.

\subsubsection{Task-Conditioned Prompt Construction}

Given an episode summary $S_e$ and a question $q$ targeting task $T_k$, we build a standardized evaluation prompt.
Each prompt contains:
\begin{enumerate}
    \item the target task metadata: \texttt{TASK\_ID}, \texttt{TASK\_NAME}, and the formal definition of $T_k$ from the task list;
    \item the episode summary $S_e$ describing the mission, network state, and dialogue context;
    \item the question stem as written in the MCQ; and
    \item the four answer options A--D.
\end{enumerate}

The model is instructed to act as an expert 6G network AI agent evaluator and to respond \emph{only} with a JSON object of the form
\begin{verbatim}
{"answer": "..."}
\end{verbatim}
where the value must be one of \texttt{"A"}, \texttt{"B"}, \texttt{"C"}, or \texttt{"D"}.
Temperature is set to zero (or a very low value), and an optional seed is derived deterministically from the episode identifier and task identifier to improve reproducibility across runs.

Formally, for a model $M$ and a question instance $(S_e, q)$ targeting task $T_k$, we obtain a prediction
\begin{equation}
\hat{y} = f_M(T_k, S_e, q) \in \{\text{A},\text{B},\text{C},\text{D}\},
\end{equation}
where $f_M$ denotes the behavior of $M$ under the standardized system and user prompts.

\subsubsection{Answer Extraction and Robust Parsing}

In practice, some models may return additional text or non-canonical JSON.
To make the evaluation robust, we first attempt to parse the response as JSON and extract a field named \texttt{"answer"}, \texttt{"choice"}, \texttt{"label"}, or \texttt{"option"} if present.
If this fails, a fallback regex search over the raw text is applied to identify a single letter in \{\text{A},\text{B},\text{C},\text{D}\}.
If no valid option can be recovered, the prediction is treated as invalid and counted as incorrect.

Each evaluated question thus yields a tuple
\begin{equation}
(\text{episode\_id}, T_k, q, y, \hat{y}, \text{meta}),
\end{equation}
where $y$ is the ground-truth option, $\hat{y}$ is the parsed model prediction, and \text{meta} includes difficulty, source turn, and raw model output for later analysis.

\subsubsection{Overall Accuracy}

Let $\mathcal{D}$ denote the full set of evaluation questions, and let each
instance $i \in \mathcal{D}$ be associated with a task
$t_i \in \{T1,\dots,T30\}$, a correct label $y_i$, and a model prediction
$\hat{y}_i$.
The primary evaluation metric is the overall multiple-choice question (MCQ)
accuracy, defined as
\begin{equation}
\mathrm{Acc}_{\text{overall}}(M)
= \frac{1}{|\mathcal{D}|}
\sum_{i \in \mathcal{D}} \mathbb{1}\left[\hat{y}_i = y_i\right],
\end{equation}
where $\mathbb{1}[\cdot]$ denotes the indicator function.

\subsubsection{Per-Task Accuracy}

To analyze semantic communication and reasoning performance at a finer
granularity, we compute per-task accuracy.
For each task $T_k$, the per-task accuracy of model $M$ is defined as
\begin{equation}
\mathrm{Acc}_{T_k}(M)
= \frac{1}{|\mathcal{D}_{T_k}|}
\sum_{i \in \mathcal{D}_{T_k}} \mathbb{1}\left[\hat{y}_i = y_i\right],
\end{equation}
where
\[
\mathcal{D}_{T_k}
= \{ i \in \mathcal{D} \mid t_i = T_k \}
\]
denotes the subset of evaluation questions labeled with task $T_k$.

\subsubsection{Group-Level Accuracy}

To capture higher-level semantic and operational capabilities, we further
aggregate per-task accuracies into group-level accuracy scores.
Let $\mathcal{G} = \{G_1,\dots,G_5\}$ denote the five evaluation groups
corresponding to
(i) intent and policy reasoning,
(ii) network slicing and resource management,
(iii) trust, security, and SLA awareness,
(iv) AI-native networking and agentic control, and
(v) distributed intelligence and emerging 6G use cases.
Each group $G_j$ is associated with a predefined subset of tasks
$\mathcal{T}_{G_j} \subseteq \{T1,\dots,T30\}$.

The group-level accuracy of model $M$ on group $G_j$ is defined as the
(unweighted) mean of the per-task accuracies within that group:
\begin{equation}
\mathrm{Acc}_{G_j}(M)
= \frac{1}{|\mathcal{T}_{G_j}|}
\sum_{T_k \in \mathcal{T}_{G_j}} \mathrm{Acc}_{T_k}(M).
\end{equation}

This aggregation ensures that each task contributes equally to its
corresponding capability dimension, independent of task-specific sample sizes,
and enables balanced comparison of model performance across heterogeneous
semantic and operational domains.
For each evaluated question, 6G-Bench records a detailed evaluation entry
containing identifiers, task metadata, difficulty, the model’s prediction,
the ground truth, and the raw model output.
These records are stored both globally and per task, enabling fine-grained
error analysis, difficulty breakdowns, and qualitative inspection of failure
modes.

\subsubsection{Stochastic pass@k Accuracy}

In addition to deterministic single-shot accuracy, we report pass@k accuracy
for a selected subset of reasoning-intensive tasks.
For each evaluation instance $i$, the model is sampled $k$ times under
stochastic decoding, yielding predictions
$\hat{y}_i^{(1)}, \dots, \hat{y}_i^{(k)}$.
The pass@k indicator for instance $i$ is defined as
\begin{equation}
\mathrm{pass@k}_i(M)
= \mathbb{1}\left[
\exists \, j \in \{1,\dots,k\} \;:\; \hat{y}_i^{(j)} = y_i
\right].
\end{equation}

Let $\mathcal{D}^{\mathrm{pass}} \subseteq \mathcal{D}$ denote the subset of
evaluation questions associated with the selected reasoning-intensive tasks.
The overall pass@k accuracy of model $M$ is then defined as
\begin{equation}
\mathrm{pass@k}_{\text{overall}}(M)
= \frac{1}{|\mathcal{D}^{\mathrm{pass}}|}
\sum_{i \in \mathcal{D}^{\mathrm{pass}}}
\mathrm{pass@k}_i(M).
\end{equation}

Similarly, the per-task pass@k accuracy for a task $T_k$ is given by
\begin{equation}
\mathrm{pass@k}_{T_k}(M)
= \frac{1}{|\mathcal{D}_{T_k}|}
\sum_{i \in \mathcal{D}_{T_k}}
\mathrm{pass@k}_i(M),
\end{equation}
where $\mathcal{D}_{T_k} \subseteq \mathcal{D}^{\mathrm{pass}}$ denotes the set
of evaluation instances labeled with task $T_k$.

By construction, pass@k is monotonically non-decreasing in $k$, with
$\mathrm{pass@1}$ equivalent to deterministic single-shot accuracy.
Larger values of $k$ therefore quantify robustness and completeness of the semantic
reasoning under stochastic decoding, rather than retry-based execution in
deployment.

\begin{table*}[t]
\centering
\small
\setlength{\tabcolsep}{4pt}
\caption{Models evaluated on 6G-Bench, ordered by release date (newest to oldest). Params are reported in billions (B). For MoE models, we give ``total (active)'' parameters; context is the maximum input length in tokens.}
\label{tab:models}
\begin{tabular}{lcccccc}
\hline
Model & Company & Category & Open? & Params (B)$^{\dagger}$ & Context & Release \\
\hline
qwen/qwen3-coder-next             & Alibaba (Qwen) & Code / agents        & Open  & 80 (3 act.)   & 262k    & 2026-02-04 \\
openai/gpt-5.2-codex              & OpenAI         & Code                 & Prop. & --            & 400k    & 2026-01-14 \\
allenai/olmo-3.1-32b-instruct     & AllenAI (Ai2)  & General chat         & Open  & 32            & 66k     & 2026-01-06 \\
openai/gpt-5.2-chat               & OpenAI         & General chat         & Prop. & --            & 128k    & 2025-12-10 \\
mistralai/mistral-small-creative  & Mistral AI     & Creative / chat      & Open & --            & 33k     & 2025-12-16 \\
mistralai/ministral-14b-2512      & Mistral AI     & General / MM         & Open  & 14            & 262k    & 2025-12-02 \\
mistralai/ministral-8b-2512       & Mistral AI     & General / MM         & Open  & 8             & 262k    & 2025-12-02 \\
anthropic/claude-haiku-4.5        & Anthropic      & General / fast       & Prop. & --            & 200k    & 2025-10-15 \\
liquid/lfm-2.2-6b                 & Liquid AI      & Edge / small         & Open & 6             & 33k     & 2025-10-20 \\
ibm-granite/granite-4.0-h-micro   & IBM            & Tools / small        & Open  & 3             & 131k    & 2025-10-20 \\
deepseek/deepseek-v3.2-exp        & DeepSeek       & General / long-ctx   & Open  & 685 (37 act.)           & 164k    & 2025-09-29 \\
nousresearch/hermes-4-70b         & Nous Research  & Reasoning (hybrid)   & Open  & 70            & 131k    & 2025-08-26 \\
tencent/hunyuan-a13b-instruct     & Tencent        & Reasoning (MoE)      & Open  & 80 (13 act.)  & 131k    & 2025-07-08 \\
meta-llama/llama-4-maverick       & Meta           & Multimodal (V+L)     & Open  & 400 (17 act.) & 1{,}048k & 2025-04-05 \\
google/gemma-3-4b-it              & Google         & Multimodal           & Open  & 4             & 96k     & 2025-03-13 \\
microsoft/phi-4                   & Microsoft      & General / SLM        & Open  & 14            & 16k     & 2025-01-10 \\
openai/gpt-4.1-nano               & OpenAI         & General / small      & Prop. & --            & 1{,}048k & 2025-04-14 \\
amazon/nova-micro-v1              & Amazon         & General / small      & Prop. & --            & 128k    & 2024-12-05 \\
qwen/qwen-2.5-7b-instruct         & Alibaba (Qwen) & General              & Open  & 7             & 33k     & 2024-10-16 \\
meta-llama/llama-3.2-3b-instruct  & Meta           & General / small      & Open  & 3             & 131k    & 2024-09-25 \\
meta-llama/llama-3.2-1b-instruct  & Meta           & Tiny general         & Open  & 1             & 60k     & 2024-09-25 \\
meta-llama/llama-3.1-8b-instruct  & Meta           & General              & Open  & 8             & 16k     & 2024-07-23 \\
\hline
\end{tabular}

\vspace{0.25em}
\footnotesize
$^{\dagger}$For Mixture-of-Experts (MoE) models, we report ``total (active)'' parameters, where active denotes the number of parameters used per token during inference.

\textbf{Abbreviations:}
MM = multimodal; 
SLM = small language model;
MoE = mixture-of-experts;
V+L = vision--language.
\end{table*}

\begin{table*}[h]
\centering
\caption{Intent \& Policy Reasoning Performance. Models are described in Table~\ref{tab:models}.}
\label{tab:intent-policy-avg}
\scriptsize
\begin{tabular}{|l|c|c|c|c|c|c|}
\hline
\textbf{Model} & $\mathrm{Acc}_{T1}$ & $\mathrm{Acc}_{T2}$ & $\mathrm{Acc}_{T3}$ & $\mathrm{Acc}_{T12}$ & $\mathrm{Acc}_{T15}$ & $\mathrm{Acc}_{G_1}$ \\\hline
qwen/qwen3-coder-next & 0.817 & \underline{\textbf{0.867}} & \underline{\textbf{0.973}} & 0.860 & 0.915 & \underline{\textbf{0.886}} \\ \hline
meta-llama/llama-4-maverick & 0.870 & 0.823 & 0.920 & 0.860 & 0.930 & 0.881 \\ \hline
mistralai/mistral-small-creative & \underline{\textbf{0.878}} & 0.805 & 0.929 & 0.897 & 0.915 & 0.885 \\ \hline
openai/gpt-5.2-chat & 0.826 & 0.832 & 0.893 & \underline{\textbf{0.916}} & 0.922 & 0.878 \\ \hline
deepseek/deepseek-v3.2 &
0.800 & 0.858 & 0.938 & 0.860 & 0.922 & 0.876 \\ \hline
mistralai/ministral-14b-2512 & 0.861 & 0.850 & 0.929 & 0.804 & 0.907 & 0.870 \\ \hline
openai/gpt-5.2-codex & 0.817 & 0.814 & 0.929 & 0.860 & 0.922 & 0.868 \\ \hline
deepseek/deepseek-v3.2-exp & 0.791 & 0.832 & 0.929 & 0.841 & 0.922 & 0.863 \\ \hline
allenai/olmo-3.1-32b-instruct & 0.843 & 0.779 & 0.920 & 0.850 & 0.915 & 0.861 \\ \hline
nousresearch/hermes-4-70b & 0.835 & 0.832 & 0.938 & 0.813 & 0.884 & 0.860 \\ \hline
anthropic/claude-haiku-4.5 & 0.783 & 0.788 & 0.866 & 0.888 & \underline{\textbf{0.938}} & 0.853 \\ \hline
mistralai/ministral-8b-2512 & 0.783 & 0.823 & 0.920 & 0.832 & 0.907 & 0.853 \\ \hline
qwen/qwen3-vl-32b-instruct & 0.852 & 0.814 & 0.875 & 0.860 & 0.860 & 0.852 \\ \hline
microsoft/phi-4 &
0.817 & 0.832 & 0.902 & 0.822 & 0.884 & 0.851 \\ \hline
qwen/qwen3-235b-a22b-2507 & 0.739 & 0.814 & 0.893 & 0.879 & 0.899 & 0.845 \\ \hline
openai/gpt-4o-mini & 0.835 & 0.814 & 0.884 & 0.813 & 0.876 & 0.844 \\ \hline
openai/gpt-5-mini & 0.791 & 0.796 & 0.839 & 0.785 & 0.907 & 0.824 \\ \hline
mistralai/ministral-3b-2512 & 0.791 & 0.735 & 0.893 & 0.776 & 0.868 & 0.813 \\ \hline
openai/gpt-4.1-nano & 0.748 & 0.735 & 0.830 & 0.738 & 0.806 & 0.771 \\ \hline
liquid/lfm-2.2-6b & 0.713 & 0.726 & 0.920 & 0.645 & 0.814 & 0.764 \\ \hline
meta-llama/llama-3.1-8b-instruct & 0.687 & 0.717 & 0.866 & 0.701 & 0.837 & 0.762 \\ \hline
meta-llama/llama-3.2-3b-instruct & 0.652 & 0.637 & 0.875 & 0.720 & 0.822 & 0.741 \\ \hline
ibm-granite/granite-4.0-h-micro & 0.661 & 0.664 & 0.839 & 0.645 & 0.791 & 0.720 \\ \hline
qwen/qwen-2.5-7b-instruct & 0.626 & 0.690 & 0.884 & 0.533 & 0.760 & 0.699 \\ \hline
amazon/nova-micro-v1 & 0.591 & 0.619 & 0.839 & 0.570 & 0.814 & 0.687 \\ \hline
google/gemma-3-4b-it & 0.670 & 0.628 & 0.786 & 0.542 & 0.589 & 0.643 \\ \hline
tencent/hunyuan-a13b-instruct & 0.443 & 0.487 & 0.652 & 0.449 & 0.667 & 0.540 \\ \hline
meta-llama/llama-3.2-1b-instruct & 0.139 & 0.150 & 0.196 & 0.131 & 0.279 & 0.179 \\ \hline
\end{tabular}
\end{table*}

\begin{table*}[h]
\centering
\caption{Network Slicing \& Resource Management Performance. Models are described in Table~\ref{tab:models}.}
\label{tab:network-slicing-avg}
\scriptsize
\begin{tabular}{|l|c|c|c|c|c|c|c|c|c|c|}
\hline
\textbf{Model} & $\mathrm{Acc}_{T4}$ & $\mathrm{Acc}_{T5}$ & $\mathrm{Acc}_{T6}$ & $\mathrm{Acc}_{T7}$ & $\mathrm{Acc}_{T8}$ & $\mathrm{Acc}_{T13}$ & $\mathrm{Acc}_{T14}$ & $\mathrm{Acc}_{T16}$ & $\mathrm{Acc}_{T29}$ & $\mathrm{Acc}_{G_2}$ \\\hline

meta-llama/llama-4-maverick &
\underline{\textbf{0.840}} & 0.763 & 0.788 & 0.795 &
\underline{\textbf{0.823}} & 0.773 & 0.841 &
\underline{\textbf{0.794}} &
\underline{\textbf{0.829}} & \underline{\textbf{0.805}} \\ \hline

qwen/qwen3-coder-next &
0.710 & \underline{\textbf{0.814}} & 0.779 & 0.778 & 0.754 & 0.758 &
0.888 & \underline{\textbf{0.794}} & 0.795 & 0.786 \\ \hline

qwen/qwen3-235b-a22b-2507 &
0.670 & 0.729 & 0.750 &
\underline{\textbf{0.880}} & 0.738 & 0.797 & 0.794 & 0.768 & 0.795 & 0.769 \\ \hline

openai/gpt-5.2-chat &
0.790 & 0.729 & 0.846 & 0.684 & 0.708 & 0.789 & 0.860 & 0.677 & 0.760 & 0.760 \\ \hline

mistralai/mistral-small-creative &
0.680 & 0.729 & 0.808 & 0.812 &
\underline{\textbf{0.823}} & 0.750 & 0.813 & 0.735 & 0.753 & 0.767 \\ \hline

deepseek/deepseek-v3.2 &
0.610 & 0.737 & 0.712 & 0.838 & 0.738 &
\underline{\textbf{0.836}} & 0.813 & 0.781 & 0.781 & 0.761 \\ \hline

mistralai/ministral-14b-2512 &
0.710 & 0.771 & 0.779 & 0.803 & 0.815 & 0.711 & 0.804 & 0.697 & 0.740 & 0.759 \\ \hline

openai/gpt-4o-mini &
0.650 & \underline{\textbf{0.814}} & 0.740 & 0.744 & 0.746 & 0.781 & 0.813 & 0.761 & 0.753 & 0.757 \\ \hline

openai/gpt-5.2-codex &
0.740 & 0.737 &
\underline{\textbf{0.856}} & 0.675 & 0.692 & 0.781 &
\underline{\textbf{0.907}} & 0.684 & 0.719 & 0.754 \\ \hline

allenai/olmo-3.1-32b-instruct &
0.670 & 0.763 & 0.788 & 0.752 & 0.754 & 0.758 & 0.822 & 0.735 & 0.733 & 0.753 \\ \hline

mistralai/ministral-8b-2512 &
0.660 & 0.797 & 0.721 & 0.744 & 0.754 & 0.758 & 0.822 & 0.710 & 0.774 & 0.749 \\ \hline

deepseek/deepseek-v3.2-exp &
0.580 & 0.686 & 0.731 & 0.855 & 0.731 & 0.781 & 0.822 & 0.755 & 0.760 & 0.745 \\ \hline

anthropic/claude-haiku-4.5 &
0.720 & 0.763 & 0.788 & 0.598 & 0.715 & 0.812 & 0.822 & 0.703 & 0.753 & 0.742 \\ \hline

qwen/qwen3-vl-32b-instruct &
0.630 & 0.703 & 0.692 & 0.752 & 0.715 & 0.719 & 0.804 & 0.735 & 0.801 & 0.728 \\ \hline

microsoft/phi-4 &
0.650 & 0.729 & 0.712 & 0.701 & 0.754 & 0.727 & 0.794 & 0.690 & 0.767 & 0.725 \\ \hline

openai/gpt-5-mini &
0.730 & 0.703 & 0.779 & 0.607 & 0.669 & 0.711 & 0.850 & 0.645 & 0.740 & 0.715 \\ \hline

nousresearch/hermes-4-70b &
0.540 & 0.754 & 0.692 & 0.701 & 0.777 & 0.703 & 0.804 & 0.677 & 0.767 & 0.713 \\ \hline

openai/gpt-4.1-nano &
0.730 & 0.754 & 0.702 & 0.718 & 0.654 & 0.727 & 0.757 & 0.619 & 0.747 & 0.712 \\ \hline

mistralai/ministral-3b-2512 &
0.650 & 0.653 & 0.779 & 0.709 & 0.654 & 0.688 & 0.794 & 0.665 & 0.774 & 0.707 \\ \hline

liquid/lfm-2.2-6b &
0.520 & 0.703 & 0.644 & 0.513 & 0.654 & 0.656 & 0.692 & 0.561 & 0.699 & 0.627 \\ \hline

meta-llama/llama-3.1-8b-instruct &
0.640 & 0.686 & 0.663 & 0.632 & 0.631 & 0.648 & 0.710 & 0.645 & 0.705 & 0.662 \\ \hline

qwen/qwen-2.5-7b-instruct &
0.410 & 0.737 & 0.606 & 0.684 & 0.631 & 0.648 & 0.598 & 0.639 & 0.603 & 0.617 \\ \hline

meta-llama/llama-3.2-3b-instruct &
0.670 & 0.627 & 0.692 & 0.453 & 0.523 & 0.602 & 0.720 & 0.561 & 0.658 & 0.612 \\ \hline

ibm-granite/granite-4.0-h-micro &
0.370 & 0.746 & 0.567 & 0.667 & 0.615 & 0.609 & 0.579 & 0.555 & 0.603 & 0.590 \\ \hline

amazon/nova-micro-v1 &
0.380 & 0.636 & 0.500 & 0.658 & 0.669 & 0.594 & 0.570 & 0.523 & 0.562 & 0.566 \\ \hline

google/gemma-3-4b-it &
0.510 & 0.508 & 0.500 & 0.564 & 0.623 & 0.508 & 0.449 & 0.458 & 0.582 & 0.522 \\ \hline

tencent/hunyuan-a13b-instruct &
0.440 & 0.534 & 0.404 & 0.479 & 0.492 & 0.555 & 0.449 & 0.432 & 0.541 & 0.482 \\ \hline

meta-llama/llama-3.2-1b-instruct &
0.240 & 0.390 & 0.202 & 0.265 & 0.300 & 0.391 & 0.234 & 0.135 & 0.308 & 0.274 \\ \hline

\end{tabular}
\end{table*}

\begin{table*}[h]
\centering
\caption{Trust, Security \& SLA Awareness Performance. Models are described in Table~\ref{tab:models}.}
\label{tab:trust-security-avg}
\scriptsize
\begin{tabular}{|l|c|c|c|c|c|}
\hline
\textbf{Model} & $\mathrm{Acc}_{T9}$ & $\mathrm{Acc}_{T10}$ & $\mathrm{Acc}_{T26}$ & $\mathrm{Acc}_{T30}$ & $\mathrm{Acc}_{G_3}$ \\
\hline

deepseek/deepseek-v3.2 &
0.809 & \underline{\textbf{0.913}} & 0.754 & 0.876 & \underline{\textbf{0.838}} \\
\hline

deepseek/deepseek-v3.2-exp &
0.783 & 0.878 & 0.739 & 0.869 & 0.817 \\ \hline

qwen/qwen3-coder-next &
\underline{\textbf{0.835}} & 0.826 & 0.697 & \underline{\textbf{0.890}} & 0.812 \\ \hline

mistralai/mistral-small-creative &
0.774 & 0.817 & \underline{\textbf{0.789}} & 0.869 & 0.812 \\ \hline

meta-llama/llama-4-maverick &
0.783 & 0.870 & 0.718 & 0.862 & 0.808 \\ \hline

mistralai/ministral-14b-2512 &
0.774 & 0.809 & 0.768 & 0.793 & 0.786 \\ \hline

nousresearch/hermes-4-70b &
0.670 & 0.843 & 0.732 & 0.855 & 0.775 \\ \hline

openai/gpt-5.2-codex &
0.696 & 0.809 & 0.690 & 0.848 & 0.761 \\ \hline

anthropic/claude-haiku-4.5 &
0.670 & 0.861 & 0.676 & 0.800 & 0.752 \\ \hline

openai/gpt-5.2-chat &
0.687 & 0.826 & 0.690 & \underline{\textbf{0.890}} & 0.773 \\ \hline

qwen/qwen3-vl-32b-instruct &
0.757 & 0.809 & 0.725 & 0.800 & 0.773 \\ \hline

allenai/olmo-3.1-32b-instruct &
0.739 & 0.774 & 0.739 & 0.828 & 0.770 \\ \hline

mistralai/ministral-8b-2512 &
0.765 & 0.791 & 0.683 & 0.828 & 0.767 \\ \hline

openai/gpt-4o-mini &
0.722 & 0.791 & 0.711 & 0.828 & 0.763 \\ \hline

microsoft/phi-4 &
0.600 & 0.861 & 0.732 & 0.855 & 0.762 \\
\hline

openai/gpt-5-mini &
0.652 & 0.713 & 0.662 & 0.786 & 0.703 \\ \hline

openai/gpt-4.1-nano &
0.643 & 0.783 & 0.606 & 0.766 & 0.700 \\ \hline

mistralai/ministral-3b-2512 &
0.670 & 0.678 & 0.648 & 0.717 & 0.678 \\ \hline

meta-llama/llama-3.1-8b-instruct &
0.539 & 0.635 & 0.613 & 0.779 & 0.642 \\ \hline

amazon/nova-micro-v1 &
0.574 & 0.609 & 0.556 & 0.710 & 0.612 \\ \hline

liquid/lfm-2.2-6b &
0.435 & 0.704 & 0.514 & 0.786 & 0.610 \\ \hline

ibm-granite/granite-4.0-h-micro &
0.548 & 0.635 & 0.570 & 0.683 & 0.609 \\ \hline

qwen/qwen-2.5-7b-instruct &
0.565 & 0.591 & 0.606 & 0.655 & 0.604 \\ \hline

google/gemma-3-4b-it &
0.478 & 0.522 & 0.542 & 0.634 & 0.544 \\ \hline

tencent/hunyuan-a13b-instruct &
0.461 & 0.591 & 0.514 & 0.641 & 0.552 \\ \hline

meta-llama/llama-3.2-3b-instruct &
0.539 & 0.435 & 0.500 & 0.621 & 0.524 \\ \hline

meta-llama/llama-3.2-1b-instruct &
0.200 & 0.243 & 0.113 & 0.448 & 0.251 \\ \hline

\end{tabular}
\end{table*}

\begin{table*}[h]
\centering
\caption{AI-Native Networking \& Agentic Control Performance. Models are described in Table~\ref{tab:models}.}
\label{tab:ai-native-avg}
\scriptsize
\begin{tabular}{|l|c|c|c|c|c|c|c|}
\hline
\textbf{Model} & $\mathrm{Acc}_{T11}$ & $\mathrm{Acc}_{T17}$ & $\mathrm{Acc}_{T18}$ & $\mathrm{Acc}_{T19}$ & $\mathrm{Acc}_{T20}$ & $\mathrm{Acc}_{T27}$ & $\mathrm{Acc}_{G_4}$ \\
\hline

meta-llama/llama-4-maverick &
0.950 & \underline{\textbf{0.877}} & \underline{\textbf{0.885}} & \underline{\textbf{0.922}} & 0.771 & 0.726 & \underline{\textbf{0.855}} \\ \hline

openai/gpt-5.2-codex &
\underline{\textbf{0.958}} & 0.812 & 0.783 & 0.876 & \underline{\textbf{0.800}} & \underline{\textbf{0.811}} & 0.840 \\ \hline

openai/gpt-5.2-chat &
0.941 & 0.818 & 0.771 & 0.884 & \underline{\textbf{0.800}} & \underline{\textbf{0.811}} & 0.838 \\ \hline

qwen/qwen3-coder-next &
\underline{\textbf{0.958}} & 0.825 & 0.783 & 0.899 & 0.781 & 0.783 & 0.838 \\ \hline

mistralai/mistral-small-creative &
0.950 & 0.818 & 0.822 & 0.884 & 0.781 & 0.726 & 0.830 \\ \hline

allenai/olmo-3.1-32b-instruct &
0.916 & 0.812 & 0.815 & 0.845 & 0.771 & 0.708 & 0.811 \\ \hline

deepseek/deepseek-v3.2-exp &
0.908 & 0.838 & 0.828 & 0.853 & 0.686 & 0.726 & 0.807 \\ \hline

qwen/qwen3-235b-a22b-2507 &
0.882 & 0.825 & 0.783 & 0.814 & 0.724 & 0.764 & 0.799 \\ \hline

deepseek/deepseek-v3.2 &
0.916 & 0.864 & 0.803 & 0.814 & 0.695 & 0.698 & 0.798 \\
\hline

microsoft/phi-4 &
0.891 & 0.805 & 0.783 & 0.814 & 0.733 & 0.745 & 0.795 \\
\hline

openai/gpt-4o-mini &
0.908 & 0.825 & 0.815 & 0.806 & 0.724 & 0.679 & 0.793 \\ \hline

nousresearch/hermes-4-70b &
0.941 & 0.838 & 0.796 & 0.798 & 0.657 & 0.726 & 0.793 \\ \hline

anthropic/claude-haiku-4.5 &
0.916 & 0.812 & 0.771 & 0.829 & 0.724 & 0.689 & 0.790 \\ \hline

mistralai/ministral-14b-2512 &
0.908 & 0.773 & 0.777 & 0.829 & 0.762 & 0.670 & 0.787 \\ \hline

qwen/qwen3-vl-32b-instruct &
0.924 & 0.779 & 0.777 & 0.822 & 0.676 & 0.708 & 0.781 \\ \hline

openai/gpt-4.1-nano &
0.891 & 0.792 & 0.739 & 0.814 & 0.714 & 0.642 & 0.765 \\ \hline

openai/gpt-5-mini &
0.882 & 0.747 & 0.713 & 0.837 & 0.752 & 0.717 & 0.775 \\ \hline

mistralai/ministral-8b-2512 &
0.916 & 0.779 & 0.764 & 0.814 & 0.695 & 0.670 & 0.773 \\ \hline

google/gemma-3-4b-it &
0.857 & 0.695 & 0.605 & 0.581 & 0.590 & 0.472 & 0.633 \\ \hline

mistralai/ministral-3b-2512 &
0.899 & 0.695 & 0.650 & 0.860 & 0.743 & 0.613 & 0.743 \\ \hline

meta-llama/llama-3.1-8b-instruct &
0.899 & 0.760 & 0.701 & 0.713 & 0.714 & 0.594 & 0.730 \\ \hline

liquid/lfm-2.2-6b &
0.874 & 0.701 & 0.688 & 0.752 & 0.629 & 0.623 & 0.711 \\ \hline

qwen/qwen-2.5-7b-instruct &
0.824 & 0.682 & 0.675 & 0.628 & 0.619 & 0.528 & 0.659 \\ \hline

meta-llama/llama-3.2-3b-instruct &
0.908 & 0.591 & 0.554 & 0.690 & 0.648 & 0.528 & 0.653 \\ \hline

ibm-granite/granite-4.0-h-micro &
0.824 & 0.669 & 0.650 & 0.543 & 0.552 & 0.566 & 0.634 \\ \hline

amazon/nova-micro-v1 &
0.824 & 0.656 & 0.624 & 0.527 & 0.514 & 0.528 & 0.612 \\ \hline

tencent/hunyuan-a13b-instruct &
0.664 & 0.714 & 0.675 & 0.527 & 0.514 & 0.462 & 0.593 \\ \hline

qwen/qwen-2.5-7b-instruct &
0.561 & 0.528 & 0.553 & 0.703 & 0.554 & 0.579 & 0.580 \\ \hline

meta-llama/llama-3.2-1b-instruct &
0.017 & 0.299 & 0.459 & 0.054 & 0.324 & 0.255 & 0.235 \\ \hline

\end{tabular}
\end{table*}

\begin{table*}[h]
\centering
\scriptsize
\caption{Distributed Intelligence \& Emerging 6G Use Case Performance. Models are described in Table~\ref{tab:models}.}
\label{tab:distributed-intel-avg}
\begin{tabular}{|l|c|c|c|c|c|c|c|}
\hline
\textbf{Model} & $\mathrm{Acc}_{T21}$ & $\mathrm{Acc}_{T22}$ & $\mathrm{Acc}_{T23}$ & $\mathrm{Acc}_{T24}$ & $\mathrm{Acc}_{T25}$ & $\mathrm{Acc}_{T28}$ & $\mathrm{Acc}_{G_5}$ \\ \hline

meta-llama/llama-4-maverick &
\underline{\textbf{0.780}} & 0.757 & 0.833 & 0.899 & \underline{\textbf{0.893}} & 0.675 & \underline{\textbf{0.806}} \\ \hline

qwen/qwen3-coder-next &
0.758 & 0.729 & 0.798 & \underline{\textbf{0.913}} & 0.884 & 0.711 & 0.799 \\ \hline

mistralai/mistral-small-creative &
0.773 & 0.771 & 0.737 & 0.884 & 0.857 & \underline{\textbf{0.728}} & 0.792 \\ \hline

qwen/qwen3-235b-a22b-2507 &
\underline{\textbf{0.780}} & 0.757 & 0.763 & 0.884 & 0.786 & 0.693 & 0.777 \\ \hline

mistralai/ministral-14b-2512 &
0.742 & 0.715 & 0.798 & 0.870 & 0.804 & 0.693 & 0.770 \\ \hline

openai/gpt-5.2-codex &
0.773 & 0.583 & 0.798 & 0.906 & 0.848 & 0.675 & 0.764 \\ \hline

openai/gpt-5.2-chat &
0.735 & 0.611 & \underline{\textbf{0.842}} & 0.870 & 0.839 & 0.675 & 0.762 \\ \hline

deepseek/deepseek-v3.2 &
0.758 & 0.764 & 0.711 & 0.855 & 0.750 & 0.684 & 0.754 \\
\hline

deepseek/deepseek-v3.2-exp &
0.705 & \underline{\textbf{0.778}} & 0.728 & 0.848 & 0.732 & 0.684 & 0.746 \\ \hline

allenai/olmo-3.1-32b-instruct &
0.720 & 0.757 & 0.667 & 0.841 & 0.821 & 0.667 & 0.746 \\ \hline

mistralai/ministral-8b-2512 &
0.720 & 0.674 & 0.693 & 0.855 & 0.839 & 0.684 & 0.744 \\ \hline

qwen/qwen3-vl-32b-instruct &
0.689 & 0.757 & 0.658 & 0.804 & 0.768 & 0.711 & 0.731 \\ \hline

openai/gpt-4o-mini &
0.659 & 0.708 & 0.772 & 0.797 & 0.768 & 0.614 & 0.720 \\ \hline

mistralai/ministral-3b-2512 &
0.689 & 0.569 & 0.763 & 0.826 & 0.812 & 0.640 & 0.717 \\ \hline

anthropic/claude-haiku-4.5 &
0.705 & 0.604 & 0.675 & 0.848 & 0.786 & 0.632 & 0.708 \\ \hline

openai/gpt-5-mini &
0.727 & 0.583 & 0.772 & 0.819 & 0.741 & 0.596 & 0.706 \\ \hline

nousresearch/hermes-4-70b &
0.674 & 0.722 & 0.623 & 0.833 & 0.714 & 0.649 & 0.703 \\ \hline

microsoft/phi-4 &
0.591 & 0.743 & 0.614 & 0.819 & 0.759 & 0.614 & 0.690 \\
\hline

openai/gpt-4.1-nano &
0.644 & 0.667 & 0.675 & 0.833 & 0.705 & 0.605 & 0.688 \\ \hline

meta-llama/llama-3.1-8b-instruct &
0.591 & 0.604 & 0.649 & 0.768 & 0.652 & 0.509 & 0.629 \\ \hline

liquid/lfm-2.2-6b &
0.508 & 0.646 & 0.579 & 0.674 & 0.670 & 0.509 & 0.598 \\ \hline

meta-llama/llama-3.2-3b-instruct &
0.523 & 0.556 & 0.596 & 0.688 & 0.652 & 0.535 & 0.592 \\ \hline

qwen/qwen-2.5-7b-instruct &
0.561 & 0.528 & 0.553 & 0.703 & 0.554 & 0.579 & 0.580 \\ \hline

ibm-granite/granite-4.0-h-micro &
0.530 & 0.521 & 0.474 & 0.623 & 0.589 & 0.421 & 0.526 \\ \hline

amazon/nova-micro-v1 &
0.568 & 0.535 & 0.447 & 0.587 & 0.509 & 0.447 & 0.516 \\ \hline

google/gemma-3-4b-it &
0.485 & 0.500 & 0.456 & 0.529 & 0.482 & 0.404 & 0.476 \\ \hline

tencent/hunyuan-a13b-instruct &
0.424 & 0.451 & 0.386 & 0.442 & 0.384 & 0.325 & 0.402 \\ \hline

meta-llama/llama-3.2-1b-instruct &
0.265 & 0.222 & 0.184 & 0.051 & 0.152 & 0.079 & 0.159 \\ \hline
\end{tabular}
\end{table*}

\begin{figure*}[t]
    \centering

    % ---- Large models ----
    \begin{subfigure}[t]{0.48\linewidth}
        \centering
         \includegraphics[width=0.8\linewidth]{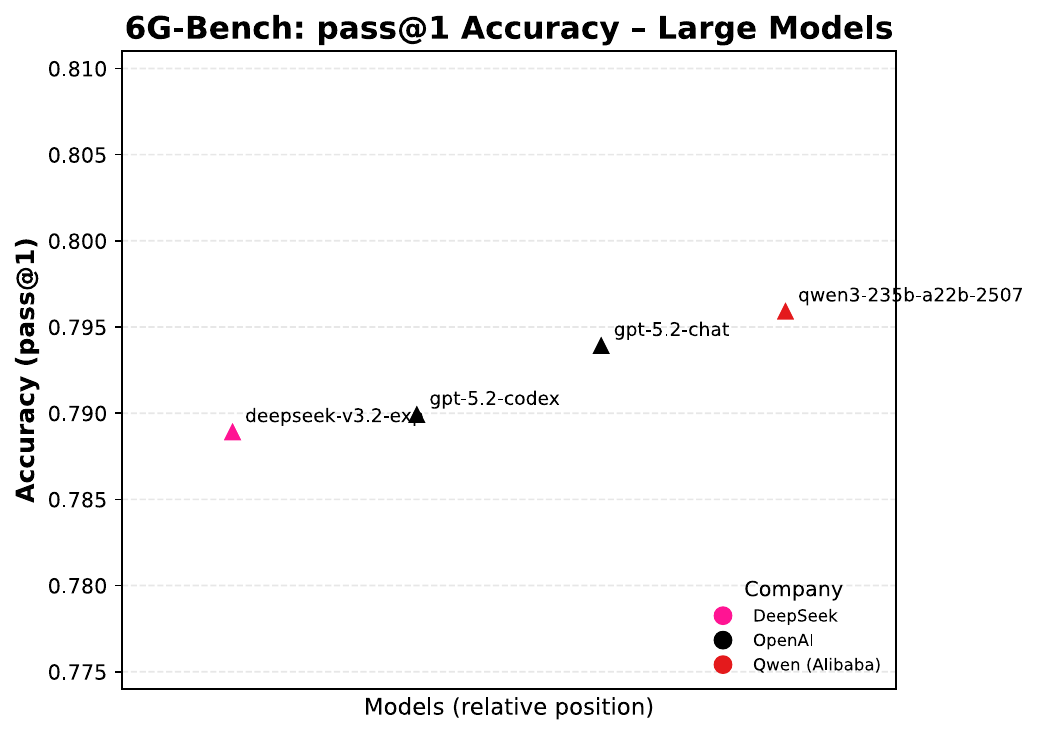}
        \caption{Large models}
        \label{fig:6g-large}
    \end{subfigure}
    \hfill
    % ---- Medium models ----
    \begin{subfigure}[t]{0.48\linewidth}
        \centering
         \includegraphics[width=0.8\linewidth]{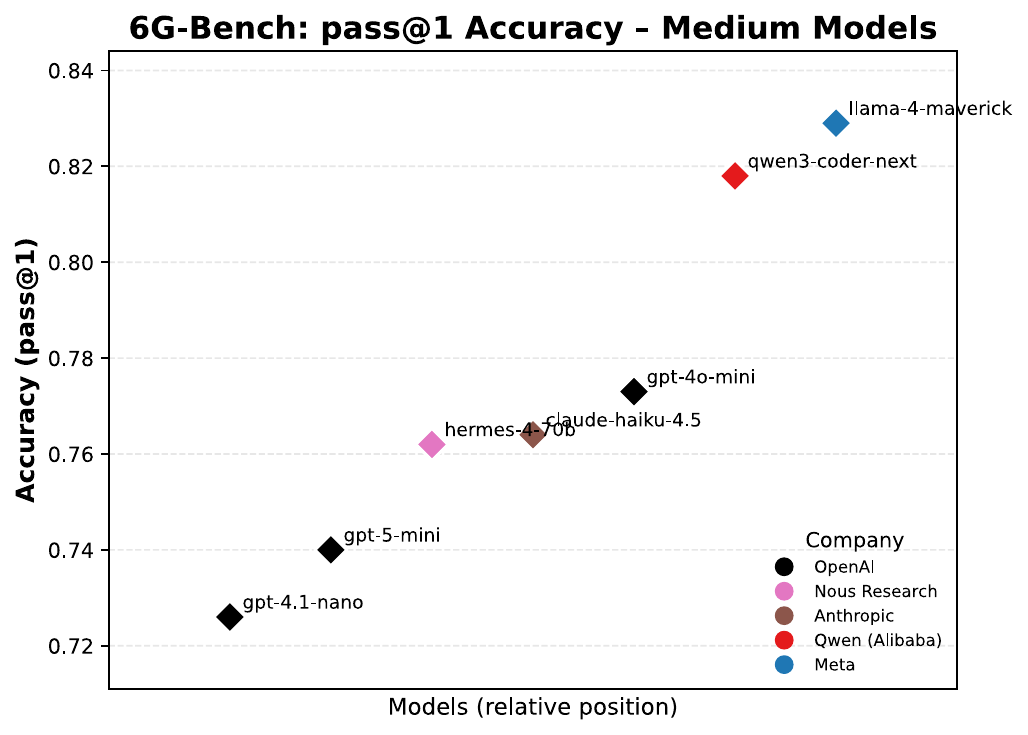}
        \caption{Medium models}
        \label{fig:6g-medium}
    \end{subfigure}

    \vspace{0.5em}

    % ---- Small models ----
    \begin{subfigure}[t]{0.48\linewidth}
        \centering
         \includegraphics[width=0.8\linewidth]{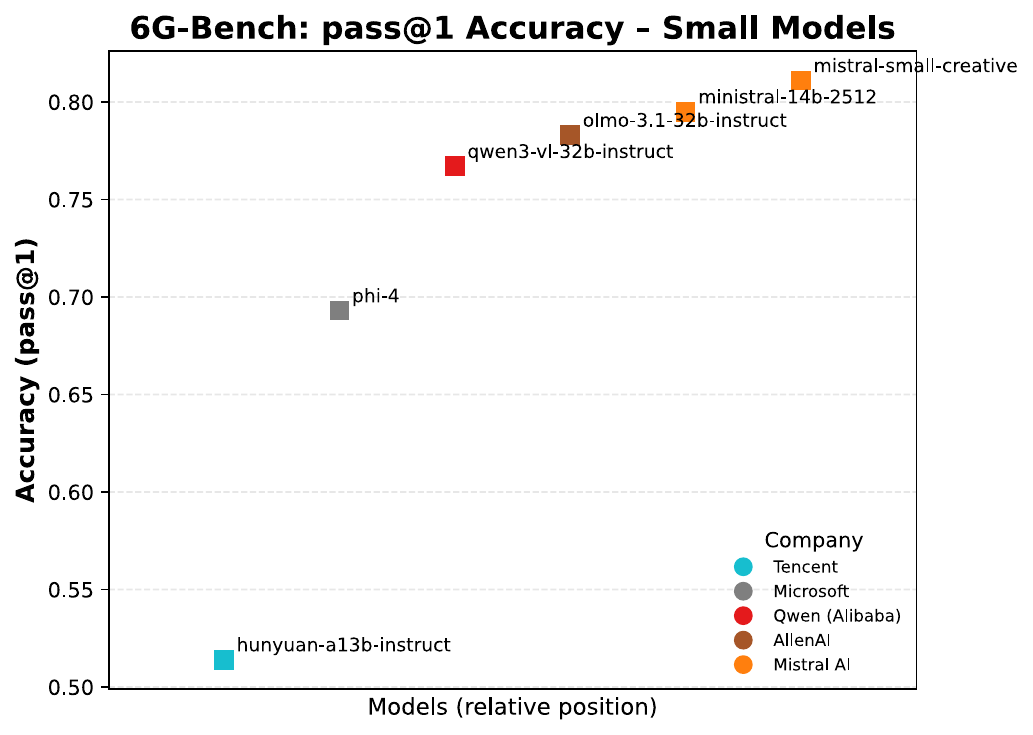}
        \caption{Small models}
        \label{fig:6g-small}
    \end{subfigure}
    \hfill
    % ---- Tiny models ----
    \begin{subfigure}[t]{0.48\linewidth}
        \centering
         \includegraphics[width=0.8\linewidth]{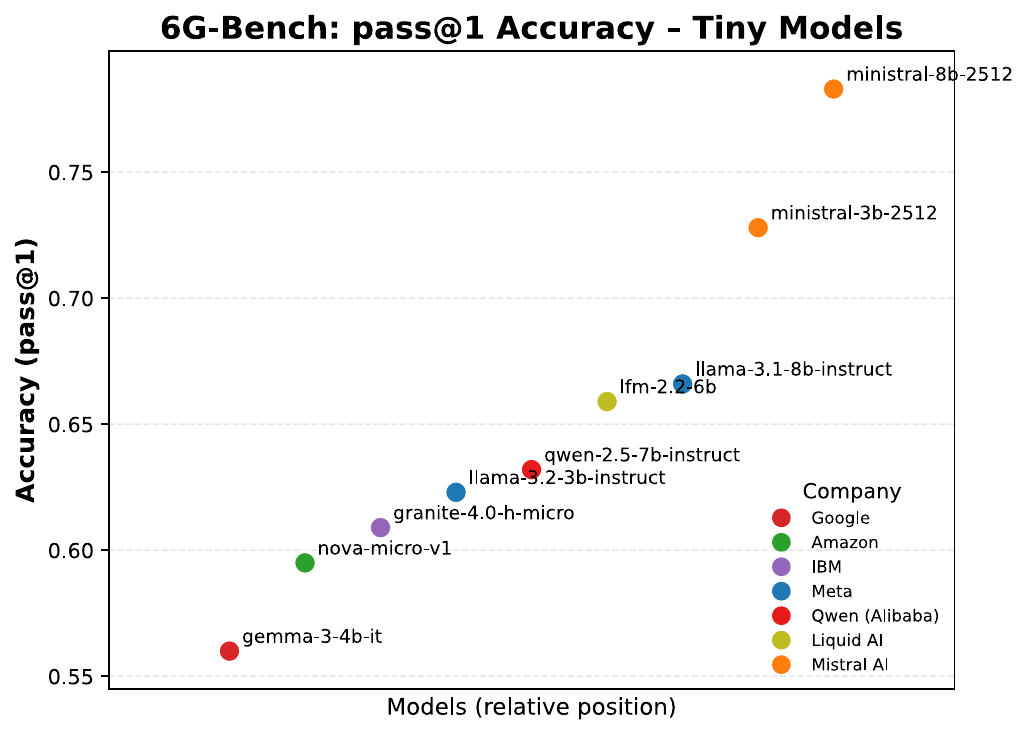}
        \caption{Tiny models}
        \label{fig:6g-tiny}
    \end{subfigure}

    \caption{
        6G-Bench pass@1 accuracy by model size category.
        Each subfigure shows pass@1 accuracy for models within a single parameter scale:
        (a) Large, (b) Medium, (c) Small, and (d) Tiny.
    }
    \label{fig:6g-bench-by-size}
\end{figure*}

\begin{table}[t]
\centering
\scriptsize
\caption{Overall pass@k performance on 6G-Bench. pass@k is reported only on selected reasoning-intensive tasks, while pass@1 corresponds to overall single-shot accuracy. Models are described in Table~\ref{tab:models}.}
\label{tab:overall_passk}
\begin{tabular}{lccc}
\hline
Model & pass@1 (Acc@1) & pass@3 & pass@5 \\
\hline
qwen/qwen3-coder-next             & 0.818 & 0.882 & \underline{\textbf{0.916}} \\
mistralai/ministral-14b-2512      & 0.795 & 0.883 & 0.915 \\
openai/gpt-5.2-codex              & 0.790 & 0.835 & 0.853 \\
deepseek/deepseek-v3.2-exp        & 0.789 & \underline{\textbf{0.888}} & 0.908 \\
nousresearch/hermes-4-70b         & 0.762 & 0.843 & 0.875 \\
anthropic/claude-haiku-4.5        & 0.764 & 0.835 & 0.867 \\
openai/gpt-5.2-chat               & 0.794 & 0.845 & 0.863 \\
meta-llama/llama-4-maverick       & \underline{\textbf{0.829}} & 0.852 & 0.859 \\
mistralai/ministral-8b-2512       & 0.783 & 0.824 & 0.852 \\
mistralai/mistral-small-creative  & 0.811 & 0.841 & 0.849 \\
allenai/olmo-3.1-32b-instruct     & 0.783 & 0.828 & 0.834 \\
meta-llama/llama-3.1-8b-instruct  & 0.666 & 0.792 & 0.831 \\
microsoft/phi-4                   & 0.693 & 0.798 & 0.821 \\
openai/gpt-4.1-nano               & 0.726 & 0.769 & 0.790 \\
meta-llama/llama-3.2-3b-instruct  & 0.623 & 0.739 & 0.789 \\
tencent/hunyuan-a13b-instruct     & 0.514 & 0.655 & 0.711 \\
amazon/nova-micro-v1              & 0.595 & 0.666 & 0.690 \\
liquid/lfm-2.2-6b                 & 0.659 & 0.685 & 0.695 \\
qwen/qwen-2.5-7b-instruct         & 0.632 & 0.662 & 0.679 \\
ibm-granite/granite-4.0-h-micro   & 0.609 & 0.613 & 0.614 \\
google/gemma-3-4b-it              & 0.560 & 0.585 & 0.588 \\
meta-llama/llama-3.2-1b-instruct  & 0.228 & 0.206 & 0.206 \\
\hline
\end{tabular}
\end{table}

\begin{table*}[t]
\centering
\caption{pass@k performance per selected reasoning-intensive task. pass@1 corresponds to deterministic single-shot accuracy, while pass@3 and pass@5 measure robustness under stochastic decoding. Models are described in Table~\ref{tab:models}.}
\label{tab:task_centric_passk}
\resizebox{\textwidth}{!}{
\begin{tabular}{|l|ccc|ccc|ccc|ccc|ccc|ccc|ccc|}
\hline
 & \multicolumn{21}{|c|}{Task ID} \\ \hline
Model
 & \multicolumn{3}{|c|}{T2}
 & \multicolumn{3}{|c|}{T9}
 & \multicolumn{3}{|c|}{T12}
 & \multicolumn{3}{|c|}{T19}
 & \multicolumn{3}{|c|}{T20}
 & \multicolumn{3}{|c|}{T26}
 & \multicolumn{3}{|c|}{T30} \\ 
 & pass@1 & pass@3 & pass@5
 & pass@1 & pass@3 & pass@5
 & pass@1 & pass@3 & pass@5
 & pass@1 & pass@3 & pass@5
 & pass@1 & pass@3 & pass@5
 & pass@1 & pass@3 & pass@5
 & pass@1 & pass@3 & pass@5 \\
\hline
mistralai/ministral-14b-2512
 & 0.858 & 0.885 & 0.912
 & 0.757 & 0.861 & 0.896
 & 0.832 & 0.907 & 0.925
 & 0.837 & 0.907 & 0.946
 & 0.724 & 0.857 & 0.905
 & 0.754 & 0.873 & 0.901
 & 0.821 & 0.890 & 0.917 \\
\hline
openai/gpt-5.2-chat
 & 0.832 & 0.841 & 0.885
 & 0.687 & 0.757 & 0.757
 & 0.916 & 0.925 & 0.925
 & 0.884 & 0.938 & 0.953
 & 0.800 & 0.838 & 0.838
 & 0.690 & 0.732 & 0.775
 & 0.890 & 0.890 & 0.910 \\
 \hline

 deepseek/deepseek-v3.2-exp
 & 0.832 & 0.894 & 0.912
 & 0.783 & 0.878 & 0.896
 & 0.841 & 0.897 & 0.935
 & 0.853 & 0.930 & 0.946
 & 0.686 & 0.838 & 0.838
 & 0.739 & 0.838 & 0.845
 & 0.869 & 0.931 & 0.972 \\
 \hline
openai/gpt-5.2-codex
 & 0.814 & 0.858 & 0.876
 & 0.696 & 0.696 & 0.748
 & 0.860 & 0.935 & 0.944
 & 0.876 & 0.915 & 0.915
 & 0.800 & 0.829 & 0.838
 & 0.690 & 0.739 & 0.754
 & 0.848 & 0.883 & 0.903 \\
  \hline
mistralai/mistral-small-creative
 & 0.805 & 0.823 & 0.823
 & 0.774 & 0.809 & 0.809
 & 0.897 & 0.897 & 0.907
 & 0.884 & 0.884 & 0.899
 & 0.781 & 0.790 & 0.810
 & 0.789 & 0.796 & 0.803
 & 0.869 & 0.883 & 0.890 \\
\hline
allenai/olmo-3.1-32b-instruct
 & 0.779 & 0.814 & 0.823
 & 0.739 & 0.739 & 0.765
 & 0.850 & 0.916 & 0.907
 & 0.845 & 0.860 & 0.868
 & 0.771 & 0.819 & 0.829
 & 0.732 & 0.789 & 0.803
 & 0.828 & 0.862 & 0.848 \\
\hline
nousresearch/hermes-4-70b
 & 0.832 & 0.885 & 0.894
 & 0.670 & 0.783 & 0.774
 & 0.813 & 0.860 & 0.907
 & 0.798 & 0.853 & 0.899
 & 0.657 & 0.743 & 0.829
 & 0.732 & 0.838 & 0.866
 & 0.855 & 0.917 & 0.938 \\
\hline
anthropic/claude-haiku-4.5
 & 0.788 & 0.858 & 0.903
 & 0.670 & 0.765 & 0.783
 & 0.888 & 0.916 & 0.963
 & 0.829 & 0.907 & 0.915
 & 0.724 & 0.781 & 0.829
 & 0.676 & 0.761 & 0.817
 & 0.800 & 0.862 & 0.869 \\
\hline
meta-llama/llama-4-maverick
 & 0.823 & 0.850 & 0.850
 & 0.783 & 0.817 & 0.809
 & 0.860 & 0.897 & 0.897
 & 0.922 & 0.930 & 0.946
 & 0.771 & 0.819 & 0.819
 & 0.718 & 0.754 & 0.768
 & 0.862 & 0.897 & 0.917 \\
\hline
mistralai/ministral-8b-2512
 & 0.832 & 0.850 & 0.885
 & 0.765 & 0.809 & 0.835
 & 0.860 & 0.888 & 0.916
 & 0.814 & 0.876 & 0.907
 & 0.695 & 0.762 & 0.800
 & 0.697 & 0.746 & 0.754
 & 0.834 & 0.855 & 0.903 \\
\hline
meta-llama/llama-3.1-8b-instruct
 & 0.708 & 0.823 & 0.850
 & 0.513 & 0.696 & 0.730
 & 0.654 & 0.832 & 0.832
 & 0.721 & 0.868 & 0.876
 & 0.705 & 0.762 & 0.829
 & 0.599 & 0.718 & 0.768
 & 0.738 & 0.841 & 0.917 \\
\hline
mistralai/ministral-8b-2512
 & 0.832 & 0.850 & 0.876
 & 0.765 & 0.809 & 0.826
 & 0.869 & 0.888 & 0.916
 & 0.837 & 0.837 & 0.899
 & 0.705 & 0.771 & 0.810
 & 0.690 & 0.739 & 0.746
 & 0.841 & 0.876 & 0.897 \\
\hline
microsoft/phi-4
 & 0.673 & 0.788 & 0.832
 & 0.574 & 0.678 & 0.670
 & 0.720 & 0.850 & 0.860
 & 0.744 & 0.853 & 0.876
 & 0.629 & 0.762 & 0.800
 & 0.662 & 0.775 & 0.810
 & 0.766 & 0.862 & 0.883 \\
\hline
openai/gpt-4.1-nano
 & 0.735 & 0.805 & 0.805
 & 0.643 & 0.739 & 0.765
 & 0.738 & 0.776 & 0.785
 & 0.814 & 0.845 & 0.860
 & 0.714 & 0.752 & 0.781
 & 0.606 & 0.669 & 0.683
 & 0.766 & 0.800 & 0.848 \\
 \hline
meta-llama/llama-3.2-3b-instruct
 & 0.646 & 0.770 & 0.823
 & 0.530 & 0.617 & 0.713
 & 0.720 & 0.832 & 0.860
 & 0.698 & 0.829 & 0.845
 & 0.629 & 0.752 & 0.829
 & 0.493 & 0.669 & 0.690
 & 0.634 & 0.724 & 0.786 \\
\hline
tencent/hunyuan-a13b-instruct
 & 0.487 & 0.611 & 0.664
 & 0.470 & 0.643 & 0.661
 & 0.449 & 0.645 & 0.757
 & 0.527 & 0.667 & 0.690
 & 0.505 & 0.571 & 0.705
 & 0.507 & 0.704 & 0.732
 & 0.634 & 0.710 & 0.759 \\
\hline
amazon/nova-micro-v1
 & 0.619 & 0.673 & 0.735
 & 0.574 & 0.626 & 0.670
 & 0.570 & 0.664 & 0.682
 & 0.527 & 0.605 & 0.659
 & 0.514 & 0.657 & 0.648
 & 0.556 & 0.676 & 0.655
 & 0.710 & 0.745 & 0.772 \\
\hline
liquid/lfm-2.2-6b
 & 0.726 & 0.788 & 0.796
 & 0.435 & 0.470 & 0.470
 & 0.645 & 0.682 & 0.701
 & 0.752 & 0.767 & 0.775
 & 0.629 & 0.667 & 0.686
 & 0.514 & 0.599 & 0.599
 & 0.786 & 0.800 & 0.821 \\
 \hline
 qwen/qwen-2.5-7b-instruct
 & 0.690 & 0.735 & 0.752
 & 0.565 & 0.600 & 0.617
 & 0.533 & 0.551 & 0.570
 & 0.628 & 0.682 & 0.690
 & 0.619 & 0.676 & 0.714
 & 0.606 & 0.641 & 0.662
 & 0.655 & 0.731 & 0.731 \\
 \hline
ibm-granite/granite-4.0-h-micro
 & 0.646 & 0.655 & 0.646
 & 0.557 & 0.591 & 0.583
 & 0.654 & 0.654 & 0.664
 & 0.527 & 0.550 & 0.550
 & 0.552 & 0.571 & 0.590
 & 0.577 & 0.577 & 0.577
 & 0.676 & 0.690 & 0.690 \\
\hline
google/gemma-3-4b-it
 & 0.628 & 0.655 & 0.646
 & 0.478 & 0.496 & 0.487
 & 0.542 & 0.542 & 0.570
 & 0.581 & 0.581 & 0.581
 & 0.590 & 0.600 & 0.610
 & 0.542 & 0.563 & 0.549
 & 0.634 & 0.648 & 0.662 \\
\hline
meta-llama/llama-3.2-1b-instruct
 & 0.150 & 0.150 & 0.150
 & 0.200 & 0.200 & 0.200
 & 0.131 & 0.131 & 0.131
 & 0.054 & 0.054 & 0.054
 & 0.324 & 0.324 & 0.324
 & 0.113 & 0.113 & 0.113
 & 0.448 & 0.448 & 0.448 \\
\hline
\end{tabular}
}
\end{table*}

\section{Performance Evaluation}
\label{sec:perf}

This section evaluates the behavior of state-of-the-art foundation models on 6G-Bench, with a focus on deployment-relevant semantics for AI-native 6G systems. The analysis proceeds in three steps. First, the benchmark setup is described, including the construction and validation of a 10{,}000-question MCQ pool derived from $\alpha^3$-Bench \cite{ferrag2026alpha} and the selection of evaluated foundation models. Second, task- and group-level accuracies are reported across intent and policy reasoning, network slicing and resource management, trust and security awareness, AI-native networking and agentic control, and distributed intelligence use cases. Third, robustness under stochastic decoding (pass@k) and the impact of model scale on deterministic pass@1 accuracy are examined, highlighting the trade-offs between semantic reliability, computational efficiency, and deployability in AI-native 6G environments.

\subsection{Benchmark Setup}

\subsubsection{Question Generation and Task Coverage}

We generated a total of 10{,}000 multiple-choice questions (MCQs) to construct 6G-Bench, using top-performing reasoning-oriented foundation models under a strictly task-conditioned generation protocol. Question generation is grounded in 113{,}475 scenarios from $\alpha^3$-Bench \cite{ferrag2026alpha}, which provides a large and diverse corpus of AI-native 6G operational contexts, including intent expression, policy constraints, network telemetry, trust boundaries, and multi-agent interactions. Each MCQ targets exactly one of the 30 benchmark tasks (T1--T30), with a dedicated, task-specific prompt that encodes the formal task definition, decision semantics, and expected reasoning scope. Generation is performed using a heterogeneous pool of large-scale, state-of-the-art language models, namely anthropic/claude-opus-4.6, openai/gpt-5.2-pro, and google/gemini-3-pro-preview, to reduce model-specific stylistic bias and increase semantic diversity across questions.

To ensure that the benchmark probes genuine high-level semantic reasoning rather than surface heuristics, all generated questions are constrained to the \emph{very hard} difficulty level. Each MCQ requires a minimum of four distinct reasoning steps, including the extraction of multiple numeric values from the scenario, the derivation of at least one quantitative margin or projection, explicit reasoning under uncertainty or worst-case bounds, and comparison of future regret across alternative actions over a multi-turn horizon (typically one to three turns). The questions integrate at least four numeric parameters drawn from multiple categories, such as network performance metrics (e.g., latency, jitter, throughput, edge load), mission or platform parameters (e.g., battery level, distance, speed), and policy or SLA limits. This structure enforces multi-factor decision-making consistent with AI-native 6G control scenarios, where no single metric is sufficient to determine the correct action.

Generation is explicitly balanced across tasks and capability groups to prevent over-representation of simpler or more templated decision types. Automatic deduplication and semantic similarity filtering are applied to remove exact and near-duplicate questions arising from scenario overlap or template instantiation. In addition, anti-heuristic constraints are enforced during generation to avoid trivial patterns, such as always favoring URLLC slices or optimizing instantaneous performance. Instead, each question is constructed so that all options incur quantifiable downsides, and the correct answer corresponds to the least-bad decision under worst-case future evolution. The full generation process is completed over approximately one week, yielding a diverse and balanced pool of 10{,}000 MCQs. Beyond evaluation, this complete set is explicitly intended for training and fine-tuning large language models for specialized 6G use cases, including semantic network control, intent-aware orchestration, and AI-native management agents.

\subsubsection{Human Expert Validation and Semantic Correctness}

From the 10{,}000 generated MCQs, we retain 3{,}722 questions as the final evaluation set following a rigorous two-stage validation process that combines automated filtering with human expert review. In the first stage, an automated validation script enforces strict structural and logical constraints, removing questions with formatting violations, inconsistent answer keys, insufficient numeric grounding, or residual duplication. Questions that do not meet the quantitative, temporal, or uncertainty-related requirements embedded in the generation prompt are automatically discarded at this stage.

In the second stage, domain experts with expertise in 6G architecture, semantic communications, and AI-native networking manually review the remaining questions over approximately 2 weeks. The expert validation focuses on semantic well-posedness, uniqueness, and correctness of the labeled answer under worst-case reasoning, and alignment with realistic 6G operational assumptions. Particular scrutiny is applied to tasks involving safety-, trust-, and policy-critical decisions, such as intent conflict resolution, conservative continuation under uncertainty, trust-aware offloading, inter-agent coordination, and network security detection and response. Questions exhibiting unresolved ambiguity, underspecified constraints, or multiple defensible answers under valid interpretations are refined or removed, even if they satisfy all automated checks.

The resulting curated set of 3{,}722 MCQs forms a high-confidence evaluation core for 6G-Bench and is used exclusively for all reported performance results, including pass@1 and selective pass@k metrics. By separating the full 10{,}000-question pool, which is suitable for training and fine-tuning, from the human-validated evaluation subset, we ensure methodological rigor and prevent leakage between training and evaluation. This validation pipeline ensures that benchmark performance reflects true semantic decision correctness under uncertainty and temporal evolution, rather than artifacts of prompt construction or generative bias, thereby aligning the evaluation with the reliability and accountability requirements of AI-native 6G deployment and standardization.

\subsubsection{Evaluated Foundation Models for 6G-Bench}

Table~\ref{tab:models} presents the foundation models considered in 6G-Bench, ordered by release date from July~2024 to February~2026, reflecting the rapid acceleration of model development for AI-native 6G systems. The selection spans code-specialized agentic models, general-purpose conversational models, multimodal vision--language systems, and compact models designed for efficiency. Context lengths range from 16k tokens, as observed in Phi-4 and Llama~3.1~8B, up to 1{,}048k tokens in Llama~4~Maverick and GPT-4.1~Nano. This wide spectrum is essential for benchmarking semantic communication workloads targeted by 6G-Bench, including reasoning-intensive tasks, long-horizon code synthesis, tool-assisted workflows, and multi-step semantic inference, all of which impose strong requirements on both memory and contextual reasoning capacity.

Architecturally, the models in Table~\ref{tab:models} illustrate a pronounced shift toward efficiency-oriented scaling strategies that align closely with deployment constraints in future 6G networks. Several recent models adopt Mixture-of-Experts (MoE) designs, notably Qwen3-Coder-Next with 80B total parameters and only 3B active per token, Hunyuan-A13B with 80B total and 13B active parameters, and Llama~4~Maverick with 400B total and 17B active parameters. These configurations enable frontier-level semantic reasoning and coding performance while substantially reducing per-token compute, a property that is particularly relevant for distributed inference across radio access networks, edge clouds, and core network intelligence layers. In parallel, smaller dense models such as Granite~4.0~H~Micro (3B), LFM~2.2~6B, and Llama~3.2~1B represent a complementary design point, supporting low-latency semantic processing, localized decision-making, and resource-constrained deployments at the network edge.

The table further highlights the strategic role of openness in the evolving AI-native 6G ecosystem. A substantial portion of the evaluated models are open-weight, including those released by Alibaba (Qwen), Meta (Llama), Mistral AI, Google (Gemma), IBM, Microsoft, Tencent, DeepSeek, and AllenAI, facilitating reproducibility, fine-tuning, and integration into experimental and pre-standard 6G semantic frameworks. Proprietary models from OpenAI, Anthropic, and Amazon are also included due to their maturity and demonstrated capability in coding, reasoning, and conversational tasks, serving as reference points for upper-bound performance. Taken together, the models summarized in Table~\ref{tab:models} capture the key trade-offs among scale, efficiency, openness, and deployability that are central to the design and standardization of semantic, AI-native 6G systems.

\subsubsection{Rationale for Selective pass@k Evaluation}
While 6G-Bench primarily targets deterministic, single-shot decision making that reflects safety- and SLA-critical network control, we report pass@k ($k \in \{3,5\}$) only for a selected subset of tasks that inherently involve multi-constraint semantic reasoning and non-trivial decision ambiguity. Specifically, tasks such as T2 (Intent Conflict Resolution), T12 (Conservative Continuation Decision), and T9/T26 (Trust-Aware Decisions) require reconciling competing objectives, for example, mission intent versus policy, performance versus trust, or progress versus safety, where multiple plausible reasoning paths may exist before converging to a correct action. Similarly, T19/T20 (Agent Coordination and Inter-Agent Communication Management) and T30 (Network Security Detection and Response Automation) involve distributed reasoning, partial observability, and anticipation of downstream effects, which are known to induce higher variance under stochastic decoding. In contrast, tasks that involve reactive, metric-driven, or procedurally constrained decisions admit little semantic ambiguity and are therefore best evaluated solely using deterministic accuracy. Restricting pass@k to reasoning-intensive tasks thus avoids inflating performance in safety-critical control decisions, while providing a meaningful measure of robustness and reasoning completeness consistent with prior LLM reasoning benchmarks.

\subsection{Intent \& Policy Reasoning Performance}

Table~\ref{tab:intent-policy-avg} presents the performance of representative foundation models on intent- and policy-aware reasoning tasks (T1, T2, T3, T12, and T15), together with the aggregated group-level accuracy $\mathrm{Acc}_{G_1}$. Overall, strong models achieve $\mathrm{Acc}_{G_1}$ values in a relatively narrow range between $0.87$ and $0.89$, indicating that intent-centric reasoning is a competitive but not yet saturated capability. The best group-level performance is obtained by qwen/qwen3-coder-next with $\mathrm{Acc}_{G_1}=0.886$, followed closely by mistral-small-creative ($0.885$) and meta-llama/llama-4-maverick ($0.881$). This clustering at the top suggests that accurate interpretation of network intent and policy constraints increasingly depends on the quality of fine-grained reasoning rather than on model scale alone.

At the task level, distinct difficulty profiles emerge. For intent feasibility assessment (T1), mistral-small-creative achieves the highest accuracy of $0.878$, slightly outperforming llama-4-maverick ($0.870$), reflecting the structured and constraint-driven nature of feasibility reasoning. Intent conflict resolution (T2), which requires semantic reconciliation between mission intent and operator policies, remains more challenging, with qwen/qwen3-coder-next achieving the best result at $\mathrm{Acc}_{T2}=0.867$, while several frontier models remain below $0.85$. In contrast, intent drift detection (T3) yields the highest absolute accuracies across all tasks, with qwen/qwen3-coder-next reaching $\mathrm{Acc}_{T3}=0.973$ and multiple models exceeding $0.93$, indicating that detecting latent semantic inconsistencies across dialog turns is comparatively easier than resolving normative conflicts.

Tasks involving uncertainty and temporal consistency expose sharper model differences. For conservative continuation decisions under uncertainty (T12), openai/gpt-5.2-chat attains the highest accuracy at $0.916$, outperforming mistral-small-creative ($0.897$) and llama-4-maverick ($0.860$), suggesting an advantage for models with stronger risk-aware alignment. For decision consistency under replanning (T15), anthropic/claude-haiku-4.5 achieves the top score of $\mathrm{Acc}_{T15}=0.938$, marginally exceeding llama-4-maverick ($0.930$) and gpt-5.2-chat ($0.922$). Finally, smaller models exhibit a pronounced performance collapse, with gemma-3-4b reaching only $\mathrm{Acc}_{G_1}=0.643$ and llama-3.2-1b dropping to $0.179$, underscoring that intent and policy reasoning in AI-native 6G systems is fundamentally non-trivial and cannot be reliably achieved by shallow or low-capacity models.

\subsection{Network Slicing \& Resource Management Performance}

Table~\ref{tab:network-slicing-avg} presents the performance of foundation models on network slicing and resource management tasks (T4, T5, T6, T7, T8, T13, T14, T16, and T29), together with the aggregated group-level accuracy $\mathrm{Acc}_{G_2}$. Compared to intent reasoning, this capability group exhibits a wider performance spread, with top models achieving $\mathrm{Acc}_{G_2}$ values between $0.76$ and $0.81$, and the best overall result obtained by meta-llama/llama-4-maverick at $\mathrm{Acc}_{G_2}=0.805$. This is followed by qwen/qwen3-coder-next ($0.786$) and qwen/qwen3-235b ($0.769$), indicating that resource-aware reasoning in AI-native 6G systems remains more challenging and sensitive to model design choices. The broader dispersion reflects the multidimensional nature of slicing decisions, which require joint reasoning across latency, reliability, compute availability, fairness, and anticipated future demand, as emphasized in the 3GPP and ETSI AI-native management frameworks.

At the task level, complementary strengths emerge across models. For slice selection reasoning (T4), llama-4-maverick achieves the highest accuracy of $0.840$, highlighting its ability to map mission semantics to URLLC, eMBB, or hybrid slices. Slice switching decisions (T5), which require recognizing when not to react to transient degradation, are best handled by qwen/qwen3-coder-next and gpt-4o-mini, both reaching $\mathrm{Acc}_{T5}=0.814$. In contrast, slice fairness versus safety (T6) remains difficult, with the top performance of $\mathrm{Acc}_{T6}=0.856$ achieved by gpt-5.2-codex, suggesting that balancing multi-agent fairness against safety-critical constraints benefits from stronger optimization-oriented reasoning. For compute placement decisions (T7), qwen/qwen3-235b achieves the highest accuracy of $0.880$, reflecting the complexity of reasoning about onboard versus edge inference under latency and SLA constraints.

Tasks involving proactive and coordinated resource adaptation further differentiate models. For graceful degradation under edge overload (T8), both llama-4-maverick and mistral-small-creative achieve the highest score of $0.823$, indicating effective anticipation of congestion-driven SLA risks. Swarm-level slice negotiation (T13) is best handled by deepseek-v3.2 with $\mathrm{Acc}_{T13}=0.836$, underscoring the difficulty of collaborative resource reasoning across multiple agents. Scheduler reconfiguration adaptation (T14) yields the highest accuracy of $\mathrm{Acc}_{T14}=0.907$ for gpt-5.2-codex, suggesting strong temporal consistency under control-plane changes. Finally, smaller models exhibit a pronounced degradation, with gemma-3-4b achieving $\mathrm{Acc}_{G_2}=0.522$ and llama-3.2-1b collapsing to $0.274$, reinforcing that network slicing and resource management in AI-native 6G networks requires deep semantic and anticipatory reasoning that cannot be reliably supported by low-capacity models.

\subsection{Trust, Security \& SLA Awareness Performance}

Table~\ref{tab:trust-security-avg} presents the performance of foundation models on trust-, security-, and SLA-aware reasoning tasks (T9, T10, T26, and T30), together with the aggregated group-level accuracy $\mathrm{Acc}_{G_3}$. Compared to intent and slicing, this capability group exhibits a distinct performance structure, reflecting the normative and risk-sensitive nature of security decisions in AI-native 6G networks. The strongest overall performance is achieved by deepseek/deepseek-v3.2 with $\mathrm{Acc}_{G_3}=0.838$, followed by its experimental variant ($0.817$) and a cluster of leading models around $0.81$. This result highlights that trust-aware reasoning benefits from models that integrate predictive risk assessment and policy compliance, rather than purely optimizing operational metrics, in line with emerging zero-trust and SLA-governed architectures in 3GPP and ETSI.

Examining individual tasks reveals asymmetric difficulty across trust dimensions. For trust-aware offloading decisions (T9), qwen/qwen3-coder-next achieves the highest accuracy at $\mathrm{Acc}_{T9}=0.835$, indicating strong capability in rejecting performance-driven actions when authorization or trust constraints dominate. SLA violation prediction (T10) is comparatively well learned, with deepseek-v3.2 reaching a peak accuracy of $0.913$, substantially higher than most peers and suggesting effective anticipation of service degradation from early network signals. In contrast, trust-aware third-party exposure (T26) proves more challenging, with the best performance of $0.789$ achieved by mistral-small-creative, while several otherwise strong models remain below $0.74$, underscoring the complexity of reasoning over regulatory, contractual, and consent-related constraints that are not easily inferred from performance telemetry alone.

Network security detection and response automation (T30) further differentiates models along safety alignment lines. The highest accuracy, $\mathrm{Acc}_{T30}=0.890$, is jointly achieved by qwen/qwen3-coder-next and openai/gpt-5.2-chat, reflecting strong reasoning about automated mitigation, isolation, and recovery actions under adversarial conditions. However, smaller, lightweight models show pronounced degradation across all trust-related tasks, with gemma-3-4b reaching only $\mathrm{Acc}_{G_3}=0.544$ and llama-3.2-1b collapsing to $0.251$. These results confirm that trust, security, and SLA awareness constitute a fundamentally more challenging semantic reasoning layer in AI-native 6G systems, requiring models to internalize normative rules, anticipate worst-case outcomes, and prioritize safety over short-term performance gains.

\subsection{AI-Native Networking \& Agentic Control Performance}

Table~\ref{tab:ai-native-avg} presents the accuracy of evaluated models on AI-native networking and agentic control tasks (T11, T17, T18, T19, T20, and T27), summarized by the group-level metric $\mathrm{Acc}_{G_4}$. This capability cluster is particularly relevant for deployment, as it reflects whether models can participate in closed-loop control, agent coordination, and lifecycle management as envisioned in 3GPP SA6, O-RAN, and ITU work on AI-native control planes. The highest group-level performance is obtained by meta-llama/llama-4-maverick with $\mathrm{Acc}_{G_4}=0.855$, followed by openai/gpt-5.2-codex at $0.840$, openai/gpt-5.2-chat at $0.838$, and qwen/qwen3-coder-next also at $0.838$. Several additional models (mistral-small-creative, olmo-3.1-32b, deepseek-v3.2-exp, and qwen3-235b) remain in the $0.80$–$0.83$ band, indicating that a broad family of contemporary models can sustain reasonably robust agentic behavior, while still leaving a non-trivial gap to the reliability levels expected for fully autonomous deployment in operational networks.

Looking inside the group, one sees a clear hierarchy of task difficulty. Preemptive autonomy downgrade decisions (T11), which encode safety-aware control choices under anticipated degradation, are effectively handled by the best models: qwen/qwen3-coder-next and gpt-5.2-codex both reach $\mathrm{Acc}_{T11}=0.958$, with several others at or above $0.95$ (for example, mistral-small-creative and llama-4-maverick at $0.950$). By contrast, network-knowledge RAG augmentation (T17) and AI agent identity and onboarding (T18) are more discriminative. Llama-4-maverick achieves the top scores on both, with $\mathrm{Acc}_{T17}=0.877$ and $\mathrm{Acc}_{T18}=0.885$, while most competing models remain in the range $0.80$–$0.83$ for T17 and $0.77$–$0.83$ for T18. This suggests that reasoning about what telemetry and knowledge should be exposed, and how agents are registered and authenticated, is an area where only a handful of models are currently suitable for deployment as first-class control-plane participants.

The multi-agent and inter-domain aspects of AI-native networks are captured by interoperability and federation (T19), agent-to-agent communication management (T20), and agent lifecycle and management (T27). Llama-4-maverick again leads on T19 with $\mathrm{Acc}_{T19}=0.922$, while several models including gpt-5.2-codex, gpt-5.2-chat, qwen3-coder-next, and hermes-4-70b achieve values above $0.87$, indicating that the high-level semantics of federation are reasonably well captured. In contrast, agent-to-agent communication management (T20) remains challenging: the best accuracy of $\mathrm{Acc}_{T20}=0.800$ is shared by gpt-5.2-codex and gpt-5.2-chat, with many otherwise strong models dropping into the $0.69$–$0.76$ band. Lifecycle management (T27) shows a similar pattern, with gpt-5.2-codex and gpt-5.2-chat reaching $\mathrm{Acc}_{T27}=0.811$, while other top-tier models remain between $0.70$ and $0.78$. At the other end of the spectrum, compact models such as gemma-3-4b ($\mathrm{Acc}_{G_4}=0.633$), nova-micro ($0.612$), hunyuan-a13b ($0.593$), and especially llama-3.2-1b ($0.235$) fall far below the thresholds that would be acceptable for in-band control-plane deployment, underscoring that AI-native networking and agentic control semantics require both substantial capacity and dedicated alignment to meet the reliability expectations of 6G operational environments.

\begin{table*}[t]
\centering
\caption{Inference runtime and efficiency on 6G-Bench for local deployments and LLM API providers.}
\label{tab:local_runtime_efficiency}
\begin{tabular}{lclll}
\hline
\scriptsize
\textbf{Model} & \textbf{Params (B)} & \textbf{Inference Mode} & \textbf{Avg Time / Q (s)} & \textbf{Throughput (Q/s)} \\
\hline
qwen/qwen3-coder-next             & 80 (3 act.)   & Novita (API)  & 3.02 & 0.33 \\
openai/gpt-5.2-codex              & --            & OpenAI (API)  & 3.30 & 0.30 \\
allenai/olmo-3.1-32b-instruct     & 32            & Local / DeepInfra (API) & 40.42 / 1.70 & 0.025 / 0.59 \\
openai/gpt-5.2-chat               & --            & OpenAI (API)  & 3.29 & 0.30 \\
mistralai/mistral-small-creative  & --            & Mistral (API) & 2.00 & 0.50 \\
mistralai/ministral-14b-2512      & 14            & Local / Together (API) & 24.79 / 1.49 & 0.040 / 0.67 \\
mistralai/ministral-8b-2512       & 8             & Local / Mistral (API) & 15.00 / 1.84 & 0.067 / 0.54 \\
anthropic/claude-haiku-4.5        & --            & Anthropic (API) & 2.91 & 0.34 \\
liquid/lfm-2.2-6b                 & 6             & Local / Liquid (API) & 10.90 / 1.22 & 0.092 / 0.82 \\
ibm-granite/granite-4.0-h-micro   & 3             & Local / Cloudflare (API) & 9.25 / 1.16 & 0.11 / 0.86 \\
deepseek/deepseek-v3.2-exp        & 685 (37 act.) & SiliconFlow (API) & 2.91 & 0.34 \\
nousresearch/hermes-4-70b         & 70            & Local / Chutes (API) & 48.61 / 2.10 & 0.021 / 0.48 \\
tencent/hunyuan-a13b-instruct     & 80 (13 act.)  & SiliconFlow (API) & 2.24 & 0.45 \\
meta-llama/llama-4-maverick       & 400 (17 act.) & DeepInfra (API) & 2.11 & 0.48 \\
google/gemma-3-4b-it              & 4             & Local / DeepInfra (API) & 10.44 / 1.83 & 0.096 / 0.54 \\
microsoft/phi-4                   & 14            & Local / DeepInfra (API) & 24.40 / 1.43 & 0.041 / 0.70 \\
openai/gpt-4.1-nano               & --            & OpenAI (API) & 2.30 & 0.44 \\
amazon/nova-micro-v1              & --            & Amazon Bedrock (API) & 1.16 & 0.86 \\
qwen/qwen-2.5-7b-instruct         & 7             & Local / Phala (API) & 14.92 / 1.44 & 0.067 / 0.70 \\
meta-llama/llama-3.2-3b-instruct  & 3             & Local / DeepInfra (API) & 9.82 / 1.43 & 0.10 / 0.70 \\
meta-llama/llama-3.2-1b-instruct  & 1             & Local / Cloudflare (API) & 4.28 / 1.46 & 0.23 / 0.68 \\
meta-llama/llama-3.1-8b-instruct  & 8             & Local / Groq (API) & 16.32 / 1.49 & 0.061 / 0.67 \\
\hline
\end{tabular}
\vspace{0.5em}
\begin{minipage}{0.95\linewidth}
\footnotesize
\emph{Notes:} Deterministic evaluation (pass@1) over 3,722 MCQs. For local models, runtime includes prompt construction, tokenization, model inference, and answer parsing; experiments are conducted on a single NVIDIA A100 GPU (80 GB) and on a multi-GPU cluster with 8 × NVIDIA A100 GPUs (80 GB each). For LLM API providers, inference latency corresponds to the end-to-end wall-clock request time. Throughput (Q/s) is computed as the inverse of the average time per question.
\end{minipage}
\end{table*}

\subsection{Distributed Intelligence \& Emerging 6G Use Case Performance}

Table~\ref{tab:distributed-intel-avg} presents model performance on distributed intelligence and emerging 6G use cases (T21, T22, T23, T24, T25, and T28), aggregated by $\mathrm{Acc}_{G_5}$. This group captures deployment-facing scenarios where communication, sensing, learning, and control are tightly coupled across heterogeneous entities. The highest group-level accuracy is achieved by meta-llama/llama-4-maverick with $\mathrm{Acc}_{G_5}=0.806$, followed by qwen/qwen3-coder-next ($0.799$) and mistral-small-creative ($0.792$). A second tier of strong performers, including qwen3-235b ($0.777$), ministral-14b ($0.770$), and gpt-5.2-codex ($0.764$), indicates that several contemporary models can reason effectively in distributed settings, though none approach saturation. The wider spread relative to earlier groups reflects the compounded uncertainty and coordination demands inherent to multi-agent, sensing-driven 6G applications emphasized in current 3GPP, ETSI, ITU-T, and O-RAN studies.

Task-level results reveal that perception- and sensing-centric decisions are comparatively well learned, while learning orchestration and resource admission remain harder. For device–network task offload arbitration (T21), the top accuracy of $0.780$ is shared by llama-4-maverick and qwen3-235b, suggesting that models can often balance on-device execution against edge or peer offloading when latency and capability cues are explicit. Federated and collaborative learning orchestration (T22) is more discriminative: deepseek-v3.2-exp achieves the highest score at $\mathrm{Acc}_{T22}=0.778$, while several otherwise strong models drop below $0.70$, highlighting the difficulty of scheduling learning under privacy, resource, and participation constraints. Network-assisted digital twin control (T23) reaches its peak with gpt-5.2-chat at $\mathrm{Acc}_{T23}=0.842$, indicating that maintaining synchronized sensing and actuation loops is within reach for frontier models, but still sensitive to representation and temporal reasoning quality.

Sensing-enhanced decisioning for ISAC (T24) stands out as the strongest-performing task across the group, with qwen/qwen3-coder-next achieving $\mathrm{Acc}_{T24}=0.913$ and llama-4-maverick closely following at $0.899$, reflecting the relative maturity of perception-driven inference when semantics are tightly coupled to sensing inputs. Disaster and public-safety coordination (T25) also yields high accuracies for leading models, with llama-4-maverick reaching $\mathrm{Acc}_{T25}=0.893$ and qwen3-coder-next at $0.884$, suggesting promising readiness for coordinated emergency response scenarios. In contrast, training-as-a-service admission decisions (T28) remain challenging, with the best score of $0.728$ achieved by mistral-small-creative, underscoring unresolved trade-offs between resource availability, privacy exposure, and quality-of-service impact. As in other groups, compact models degrade sharply—gemma-3-4b attains $\mathrm{Acc}_{G_5}=0.476$ and llama-3.2-1b collapses to $0.159$—confirming that distributed intelligence in AI-native 6G networks requires substantial semantic capacity and anticipatory reasoning to support deployment-grade operation.

\subsection{Robustness Analysis via pass@k Evaluation}

Table~\ref{tab:overall_passk} presents the overall pass@k performance on 6G-Bench, where pass@1 corresponds to deterministic single-shot accuracy and pass@3 and pass@5 quantify robustness under stochastic decoding on selected reasoning-intensive tasks. Several deployment-relevant trends emerge. First, the strongest single-shot performance is achieved by meta-llama/llama-4-maverick with $\mathrm{pass@1}=0.829$, closely followed by qwen/qwen3-coder-next ($0.818$) and mistral-small-creative ($0.811$). However, when allowing multiple reasoning attempts, deepseek-v3.2-exp achieves the highest robustness at $\mathrm{pass@3}=0.888$, while qwen/qwen3-coder-next attains the best $\mathrm{pass@5}=0.916$. This divergence between pass@1 and pass@k rankings highlights that models with similar deterministic accuracy can differ substantially in their ability to recover correct decisions under stochastic exploration, a property that is critical for off-line analysis and decision-support scenarios but less acceptable for in-band autonomous control.

The task-centric results in Table~\ref{tab:task_centric_passk} provide deeper insight into the origins of robustness gains. For intent conflict resolution (T2), pass@5 values reach $0.912$ for deepseek-v3.2-exp and $0.912$ for ministral-14b, compared to pass@1 values around $0.83$–$0.86$, indicating that alternative reasoning paths frequently converge to the correct semantic reconciliation. Similar behavior is observed for conservative continuation under uncertainty (T12), where several models exceed $\mathrm{pass@5}=0.93$ (for example, gpt-5.2-codex at $0.944$ and claude-haiku-4.5 at $0.963$), even when pass@1 is below $0.90$. In contrast, trust-aware offloading (T9) and trust-aware third-party exposure (T26) exhibit more limited robustness gains, with pass@5 improvements typically below $0.15$, reflecting the sharper normative constraints and reduced ambiguity inherent in trust and authorization decisions.

Multi-agent and security-oriented tasks further illustrate the deployment implications of pass@k behavior. For agent interoperability and federation (T19), several models achieve very high robustness, with pass@5 reaching $0.946$ for llama-4-maverick and deepseek-v3.2-exp, suggesting that coordination semantics are often recoverable given multiple reasoning attempts. Network security detection and response automation (T30) shows the strongest overall robustness gains, with deepseek-v3.2-exp achieving $\mathrm{pass@5}=0.972$ and multiple models exceeding $0.90$, indicating that threat-response reasoning admits multiple valid inference paths before converging to a correct action. At the same time, compact models display unstable or even degraded pass@k behavior (for example, llama-3.2-1b remains at $0.448$ across all k), underscoring that robustness under stochastic decoding is not a substitute for sufficient semantic capacity. Taken together, these results justify the selective use of pass@k in 6G-Bench: while it provides valuable insight into reasoning completeness and ambiguity resolution, deterministic pass@1 accuracy remains the appropriate primary metric for safety- and SLA-critical AI-native 6G deployment.

\subsection{Model Scale Effects on pass@1 Performance}

Table~\ref{tab:overall_passk}, together with Fig.~\ref{fig:6g-bench-by-size}, presents the pass@1 accuracy of the evaluated models, grouped by parameter scale into large, medium, small, and tiny categories. A first-order observation is that pass@1 accuracy does not grow monotonically with model size. Among large models, pass@1 values cluster tightly between $0.789$ and $0.796$, with qwen3-235b-a22b-2507 achieving $0.796$, gpt-5.2-chat reaching $0.794$, gpt-5.2-codex at $0.790$, and deepseek-v3.2-exp at $0.789$. This narrow spread suggests that, at very large scales, performance on semantic and network-level reasoning saturates under deterministic decoding, and further parameter increases yield diminishing returns without corresponding gains in architectural or alignment design.

In contrast, medium-scale models exhibit the strongest overall pass@1 performance. As shown in Fig.~\ref{fig:6g-medium}, meta-llama/llama-4-maverick achieves the highest pass@1 accuracy of $0.829$, outperforming all larger models, while qwen3-coder-next follows closely at $0.818$. Other medium models such as gpt-4o-mini ($0.773$), claude-haiku-4.5 ($0.764$), and hermes-4-70b ($0.762$) form a second tier. This regime highlights a critical deployment-relevant insight: models in the medium parameter range appear to offer the best balance between capacity and controllability for AI-native 6G reasoning, aligning with standardization expectations that future network intelligence components must remain efficient, auditable, and deployable at scale rather than relying on extreme model sizes.

The degradation trend becomes more pronounced as the model size decreases. Small models (Fig.~\ref{fig:6g-small}) span a wide range, from mistral-small-creative at $0.811$ and ministral-14b-2512 at $0.798$, down to hunyuan-a13b-instruct at $0.514$. Tiny models (Fig.~\ref{fig:6g-tiny}) show another collapse, with the best-performing ministral-8b-2512 reaching $0.783$, while many others remain between $0.56$ and $0.66$, and the weakest configurations falling below $0.60$. These results indicate that while carefully designed small models can still support limited semantic reasoning, ultra-compact models lack the representational depth required for reliable intent, trust, and coordination decisions. From a 6G deployment perspective, this reinforces that AI-native control-plane intelligence will likely converge on a mid-scale model regime, where semantic reasoning accuracy, robustness, and operational feasibility intersect most effectively.

\subsection{Inference Efficiency and Throughput Analysis}

Table~\ref{tab:local_runtime_efficiency} reveals pronounced efficiency gaps between local single-GPU inference and API-hosted deployments across a wide range of model scales. Inference services provided by Amazon Bedrock, IBM Cloudflare, Liquid AI, OpenAI, Mistral, and DeepInfra consistently deliver substantially higher throughput than local execution on a single NVIDIA A100, often exceeding it by an order of magnitude. Notably, lightweight offerings such as Amazon Nova Micro v1 and IBM Granite-4.0-H Micro achieve the highest throughput (approximately 0.86~Q/s), corresponding to average latencies of approximately 1 second per query, while models served via OpenAI, Anthropic, and Mistral maintain competitive throughput (approximately 0.3--0.5~Q/s) despite substantially larger model capacities. In contrast, local inference exhibits steep scaling inefficiencies with model size: models above 30B parameters incur per-question latencies exceeding 20--40~s, resulting in throughput below 0.05~Q/s, and even smaller local models remain 6--15$\times$ slower than their API counterparts. These results indicate that provider-managed inference stacks, which leverage optimized kernels, request batching, and multi-accelerator parallelism, substantially mitigate latency and throughput bottlenecks, positioning hosted inference as the dominant solution for latency-sensitive and high-throughput evaluation workloads, while local deployments primarily trade efficiency for increased controllability and deployment autonomy.

\subsection{Implications for 6G Semantic Communication Systems}

The results in Table~\ref{tab:local_runtime_efficiency} have direct implications for the design of future 6G semantic communication systems. The substantial throughput gap between single-GPU local inference and API-hosted deployments suggests that telecom operators aiming to support latency-sensitive semantic services will need to move beyond isolated accelerators and instead provision clustered GPU infrastructures at the network edge or core. By adopting multi-GPU serving architectures similar to those employed by commercial API providers, operators can approach comparable inference efficiency while retaining full control over data locality, model configuration, and operational policies. Such an approach enables semantic processing to be integrated within the trusted telecom domain, thereby preserving data privacy and regulatory compliance, which are critical requirements in 6G applications involving user-generated content, industrial telemetry, and mission-critical communications.

\section{Conclusion}
\label{sec:conc}

In this paper, we proposed \emph{6G-Bench}, an open and standardized benchmark for evaluating semantic communication and network-level reasoning with foundation models in AI-native 6G networks, grounded in ongoing 6G and AI-agent standardization activities in 3GPP, IETF, ETSI, ITU-T, and the O-RAN Alliance. We extracted a taxonomy of 30 decision-making tasks spanning intent and policy reasoning, network slicing and resource management, trust and security awareness, AI-native networking and agentic control, and distributed intelligence, and instantiated them using 10{,}000 very-hard multiple-choice questions generated from 113{,}475 scenarios under task-conditioned prompts enforcing multi-step quantitative reasoning, uncertainty awareness, and worst-case regret minimization. After automated filtering and expert human validation, 3{,}722 questions were retained as a high-confidence evaluation set, and we evaluated 22 contemporary foundation models spanning dense and mixture-of-experts architectures, short- and long-context designs (up to 1M tokens), and both open-weight and proprietary systems. The results reveal substantial variability in semantic reasoning capability, with deterministic single-shot performance (pass@1) ranging from 0.228 to 0.829, intent and policy reasoning accuracy for leading models clustered between 0.87 and 0.89, and selective robustness on reasoning-intensive tasks reaching pass@5 = 0.916. These findings indicate that mid-scale foundation models currently offer the most favorable balance between accuracy, robustness, and deployability, while trust-aware, security-critical, and distributed intelligence tasks remain key bottlenecks. Overall, 6G-Bench provides a quantitative and reproducible foundation for assessing the readiness of foundation models as semantic reasoning layers above standardized network functions, with direct implications for AI-native 6G deployment and future standardization.

\bibliographystyle{IEEEtran}
\bibliography{bibliography}

\end{document}